\documentclass[12pt,a4paper]{article}
\usepackage[utf8]{inputenc}
\usepackage{authblk}
\usepackage{amsmath}
\usepackage{cite}
\usepackage{color}
\usepackage{soul}
\usepackage{amsfonts}
\usepackage{amssymb}
\usepackage{mathrsfs}
\RequirePackage{graphicx}
\usepackage[left=2cm, right=1cm, top=2cm]{geometry}

\DeclareMathOperator*{\sumint}{%
\mathchoice%
  {\ooalign{$\displaystyle\sum$\cr\hidewidth$\displaystyle\int$\hidewidth\cr}}
  {\ooalign{\raisebox{.14\height}{\scalebox{.7}{$\textstyle\sum$}}\cr\hidewidth$\textstyle\int$\hidewidth\cr}}
  {\ooalign{\raisebox{.2\height}{\scalebox{.6}{$\scriptstyle\sum$}}\cr$\scriptstyle\int$\cr}}
  {\ooalign{\raisebox{.2\height}{\scalebox{.6}{$\scriptstyle\sum$}}\cr$\scriptstyle\int$\cr}}
}

\begin{document}

\title{\bf HTL effective action of topologically massive\\ gluons in 3+1 dimensions}

\author[a]{Debmalya Mukhopadhyay\thanks{Electronic address: debphys.qft@gmail.com}}
\author[b]{R. Kumar\thanks{Electronic address: raviphynuc@gmail.com}}
\author[a]{Jan-e Alam\thanks{Electronic address: jane@vecc.gov.in}}
\author[a]{Sushant K. Singh\thanks{Electronic address: sushantsk@vecc.gov.in}}
\affil[a]{\small Variable Energy Cyclotron Centre, 1/AF, Bidhan Nagar, Kolkata - 700064, India}
\affil[b]{\small Department of Physics and Astrophysics, University of Delhi, New Delhi-110007, India}

\maketitle
\abstract{We construct an effective action for ``soft'' gluons by integrating out hard thermal modes of topologically massive 
vector bosons at one loop order. The gluons are equally
massive in the non-Abelian topologically massive model (TMM) due to a quadratic coupling $B\wedge F$ where a 2-form field 
$B$ is coupled quadratically with the  field strength $F$ of Yang-Mills (YM) field. The presence of infrared cut-off in 
the model can be used to get color diffusion  constant and conductivity in perturbative regime.}

\vskip .5 cm
\noindent 
{\bf PACS numbers:} 11.15.-q; 12.38.-t; 12.38.Mh
\vskip .2cm
\noindent
{\bf Keywords:} Topologically massive $B \wedge F$ theory; QCD; hard thermal loop; quark gluon plasma; thermal field theory


\section{Introduction}

Gauge theory plays a crucial role in the standard model of particle physics for the description of fundamental interactions in 
nature~\cite{sw:1967, as:68, sg:61}. The standard model is the theory that describes  three fundamental interactions 
(i.e. electromagnetic, weak and strong) among all the known particles excluding gravitational interaction. In the electroweak sector, global $SU(2)\times U(1)$ 
symmetry is spontaneously broken to global $U_{\text{em}}(1)$ symmetry. This residual symmetry is responsible for the  electromagnetic 
interaction. The mediators of the weak force, $W^{\pm}$ and $Z$ bosons, become massive via Higgs mechanism through the process of 
spontaneous symmetry breaking. The Higgs particle has been discovered in the large hadron collider (LHC)~\cite{atlas,cms}.

The strong sector in the standard model has a special characteristic which makes it to be significantly different from the 
electroweak sector. The elementary particles, quarks and gluons, which interact strongly are not found free in 
any experiment till date. The dynamics of quarks and gluons is governed by quantum chromodynamics (QCD). The confinement of 
the quarks within the hadrons is yet to be understood. Beside this, one of the other important features of the strong interaction is the asymptotic 
freedom which implies the validity of perturbative analysis of QCD interaction in high energy limit\footnote{If 
the energy of centre of momentum frame of collision be $E$, then here the high energy limit implies $E\gg m$ for any mass $m$ 
present in the interaction.}~\cite{gw:1973,pol:1973,reya:1981,sc:1973,zee:1973,frankel:1976,gross:2005,hooft:1985,QCD}. 
The asymptotic freedom also helps us to realize a deconfined state of matter in QCD known as quark-gluon plasma 
(QGP) at high density and temperature~\cite{qgp}. 

QGP is a thermal system  of deconfined  quarks and gluons. It can be created by colliding nuclei at 
ultrarelativistic energies such as Relativistic Heavy Ion Collider (RHIC)~\cite{rhicexpt} and Large Hadron Collider(LHC)~\cite{lhcexpt} energies. 
It may be formed after $1$fm/c ($\sim \Lambda_{\text{QCD}}^{-1}$) of nuclear collision.
The transport coefficients of QGP like shear viscosity,
bulk viscosity etc. can be used to characterize the QGP by studying its hydrodynamic evolution. In our present endeavor, 
we are interested in the perturbative aspects of QGP where gluon degrees of freedom dominate. Such a state can be 
created by colliding nuclei at LHC and higher RHIC energies. The study of QGP offers an opportunity to understand QCD in medium.


QGP state also provides an opportunity to investigate the  non-trivial topological configurations of gauge fields. 
The non-trivial  topological configuration localized in $(3+1)$-dimensions of space-time is known to be instanton. 
This configuration shows that the Yang-Mills (YM) theory has infinite vacua. These vacua are designated by a parameter 
$\theta$. Instanton carries a great importance in producing the chiral magnetic effect in QGP when massless
quarks are considered. This effect is a combination of electromagnetic and  chromomagnetic 
phenomena~\cite{kharzeev,warringa,Fukushima,akamatsu1}. The chiral imbalance can  help us to investigate the 
violation of parity $\mathcal{P}$ and $\mathcal{CP}$ symmetries in QCD\footnote{Here $\mathcal{C}$ designates 
charge conjugation operation.} (strong $\mathcal{CP}$ problem). 

QGP is considered often with massless gluons\footnote{Here ``massless gauge field'' implies the gauge field 
having ``bare mass'' mass at zero temperature.}. However, gluons can acquire non-zero masses (i.e. electric and magnetic masses)  at finite temperature. 
These masses were shown to be gauge invariant~\cite{braaten,kobes}.  The masses carry a 
great importance in the analysis of QGP~\cite{linde,furusawa}. Electric mass provides the Debye screening whereas 
the non-zero magnetic mass implies the validity of the application of perturbation technique in the analysis of QGP. 
Debye mass also plays a pivotal  role in the suppression of the effect of large instanton in QGP. On the other hand, 
it is shown that magnetic mass is absent in massless non-Abelian gauge theory in every loop correction~\cite{furusawa} 
and  hence it is treated in non-perturbative regime at the length scale $\sim 1/(g^2T)$ which is much below the scale 
of mean free path $\sim 1/(g^4 T)$; here $g (<1)$ is the QCD gauge coupling. It can be shown that the dynamical screening  
can prevent the infrared singularities in QED plasma, 
but this would not work for QCD plasma because the  massless gluon fields carry color charges.

In this paper, we will construct an effective Lagrangian density by integrating out the hard modes of 
topologically massive gluons (with momentum $\sim T$). This procedure has been followed to obtain a general form of HTL-effective action  
where the massless  gauge fields are present~\cite{stanis}. But we consider an effective action for massive gauge field.  The effective action, 
obtained here for massive gauge fields,  will be useful for the 
computation of the color conductivity and color diffusion constant~\cite{arnold,boodeker,selikov,heiselberg} in perturbative regime. 
At $T=0$, the  massless non-Abelian gauge field has a problem in the description of local interaction in quantum 
field theory (QFT)~\cite{stro,haag}. 
Since the Fock space of the non-Abelian gauge field has positive indefinite metric, the interactions among 
the  massless gluons violate cluster decomposition principle~\cite{stro,kugo,fischer} which is not desirable 
in a Lorentz invariant model. On the other hand, massive gluon can explain the color singlet asymptotic states 
in physical Hilbert space in QCD~\cite{kugo,chaichian} when color symmetry is not broken spontaneously. But 
the presence of mass in the pure non-Abelian gauge theory causes many other problems. For instance, the gauge 
bosons acquire longitudinal mode which violates unitarity in the scattering processes at 
high energy limit. This can be seen in any massive non-Abelian gauge theory, for example, electroweak 
sector~\cite{sg,cev,lee}. However, in this sector, these are the Higgs mediated processes which recover the unitarity 
of scattering matrix. But color symmetry is believed to be an exact symmetry in the strong sector. Hence, the Higgs mechanism 
and Proca theory cannot be taken  into consideration. We can also think of the non-Abelian St{\"u}ckelberg model, 
but it was found to be non-renormalizable~\cite{veltman:1968, veltman:1970,shizuya:1977,umezawa:1961,Ruegg:2003ps}.  
The Curci-Ferrari model contains Proca-massive gauge field and it was found to be non-unitary in spite of 
being renormalizable~\cite{cf:75,oma:82}. There was also an attempt for the dynamical generation of  mass of  YM 
field~\cite{cl:82}, but that mass vanishes in  high energy limit~\cite{cornwall:2011}.


The $(3+1)$-dimensional TMM contains a topological term: 
$mB\wedge F = \dfrac{m}{4}\,\varepsilon^{\mu\nu\rho\lambda}\,F_{\mu\nu} B_{\rho\lambda}$~\cite{balu}. 
Here $B$ is a two-form field  and $F$ is the field strength of the one-form gauge field $A$. This is topological
field theory of Schwarz-type~\cite{balu,birm}.  
This term is a key ingredient for the field 
theories which are to be independent of metric. For example, in the formulation of quantum gravity, this 
term is used for the action~\cite{baez}. In QFT, by considering the kinetic terms of  $A$ and $B$ fields, a 
model can be constructed where observables are related to the local excitations and topological invariants 
in TMM~\cite{birm,horow}. We observe that the coupling 
constant $m$  becomes the pole for the gauge field propagator when the $B$-field is integrated out.  The spin representation 
of the $B$ field is different from the $A$ field. Unlike $A$ field, the massless $B$ field has one degree of freedom whereas the massive 
$B$ field behaves like a massive one-form field in the Lorentz representation~\cite{kr:1973}. Hence, by integrating out  
either $A$ or $B$ in the TMM, we obtain an effective field theory for the  massive vector bosons.  We also see that 
the TMM is invariant under the vector gauge symmetry of $B$ field beside the vector gauge symmetry of YM 
field. The presence of infrared cut-off in the non-Abelian generalization of the TMM validates  the perturbative 
analysis in the massive quantum gauge theory.

The contents of our present endeavor are as follows. In section 2, we discuss the non-Abelian  
TMM very briefly. Section 3 deals with the various vertex rules, propagators of 
the gauge and ghost fields present in the TMM. We also show, in this section,  how the coupling constant ``$m$'' becomes the pole of the 
complete propagator of YM field. In the whole calculations, the signature of 4D Minkowski metric $\eta_{\mu\nu}$ 
is chosen as $\text{diag} (+,~-,~-,~-)$  and $\hbar = k_B = 1$  where $\hbar$  and $k_B$ are the Plank and  Boltzmann constants,  
respectively. In section 4, we estimate the thermal mass for one-form massive 
gauge field at one loop order. In this section, the hard thermal modes of one-form, two-form and ghost 
fields are integrated out at one loop order and  an effective action for soft massive gluons is obtained. 
Finally,  section 5 has been dedicated to discuss the implication of the results, obtained in this work.



\section{($3+1)$-dimensional (4D) topologically massive model}

The Lagrangian density of the model is given by~\cite{cs:1974, aab:1990, al:01}
\begin{eqnarray}
\mathscr{L}& = & -\frac{1}{4}\,F_{\mu\nu}^a F^{a\mu\nu}  + \frac{1}{12}\, \tilde{H}_{\mu\nu\lambda}^a  \tilde{H}^{a\mu\nu\lambda} 
+ \frac{m}{4}\,\varepsilon^{\mu\nu\rho\lambda}\,B_{\mu\nu}^a\, F_{\rho\lambda}^a,
\label{nonab-bf}
\end{eqnarray}
where the field strengths corresponding the Yang-Mills field $A^a_\mu$ and the two-form gauge field 
$B^a_{\mu\nu}$ are, respectively, given by
\begin{eqnarray}
F^a_{\mu\nu} & = & \partial_\mu A^a_\nu - \partial_\nu A^a_\mu + g f^{abc}\, A^b_\mu\, A^c_\nu, 
\end{eqnarray} 
and
\begin{eqnarray}
\tilde{H}^a_{\mu\nu\lambda} &=& (D_{[\mu} \,B_{\nu\lambda]})^a - g f_{abc} \,F_{[\mu\nu}^b\, C_{\lambda]}^c \nonumber\\
&=& \partial_{[\mu} \, B^a_{\nu\lambda]} + g f^{abc}\, A^b_{[\mu} \, B^c_{\nu\lambda]} - g f_{abc} \,F_{[\mu\nu}^b\, C_{\lambda]}^c,
\end{eqnarray}
where the fields $A^a_\mu$, $B^a_{\mu\nu}$ and $C^a_\mu$ are in the adjoint representation of the $SU(N)$ gauge group. 
Unlike the Abelian model (see Section 3 below), we have an extra vector field $C_\mu^a$ in this model. It is an auxiliary field~\cite{meig:1983} which  
assures the invariance of the Lagrangian density under the following transformations
\begin{eqnarray}
A_{\mu}^a \to A_{\mu}^a  \qquad B_{\mu\nu}^a \to B_{\mu\nu}^a + \left(D_{[\mu}\, \theta_{\nu]}\right)^a,
\qquad C_\mu^a \to C_\mu^a + \theta_{\mu}^a,
\end{eqnarray}
where  $\theta_{\mu}^a$ is a vector field in adjoint representation of $SU(N)$. Including the Faddeev-Popov ghost fields and Nakanishi-Lautrup 
fields corresponding to the $A_\mu^a$ and $B_{\mu\nu}^a$ fields, we get the full action~\cite{al:01} as 
\begin{eqnarray}
S = S_0 & + & \int d^4x \Big[h^af^a + \frac{\xi}{2}\, h^a h^a - h_\mu^a \big(f^{a\mu} + \partial^\mu n^a \big)\nonumber\\
&-& \beta^a \big(\partial_\mu D^\mu \beta^a - \partial_\mu(g f^{abc}\, \omega^{b \mu}\, \omega^c) \big) - \frac{\eta}{2}\, h_\mu^a \,h^{a\mu} 
+ \partial_\mu \bar{\omega}^{a\mu}\, \alpha^a - \bar{\alpha}^a\, \partial_\mu \omega^{a\mu} \nonumber\\
&-& \zeta \,\bar{\alpha}^a \,\alpha^a + \bar{\omega^a}\, \partial_\mu D^\mu \omega^a \nonumber\\
&-&  \bar{\omega}^a_\mu \big\{\partial_\nu \big(g f^{abc} \,B^{b\mu\nu}\,\omega^c \big) + \partial_\nu \big(D^{[\mu}\, \omega^{\nu]}\big)^a
+ \partial_\nu \big(g f^{abc}\, F^{b\mu\nu} \, \theta^c)\big \} \Big],
\label{nonabactn}
\end{eqnarray} 
where $S_0 (= \int d^4 x \mathscr{L})$ is the action corresponding to the Lagrangian density 
(\ref{nonab-bf}) where  $ f^a = (\partial^\mu A_\mu)^a$ and  $f^a_\mu = (\partial^\nu B_{\mu\nu})^a$. 
The parameters $\xi, \eta$ and $\zeta$ are the dimensionless gauge-fixing parameters. The auxiliary 
fields $h^a$ and $h^a_\mu$ play the role of Nakanishi-Lautrup type fields. 
Here $\omega^a$ and $\bar{\omega}^a$ are the Faddeev-Popov ghost and anti-ghost fields (with ghost number 
$+1$ and $-1$, respectively) corresponding to vector gauge field $A^a_\mu$.  The Lorentz vector ghost fields  $(\bar{\omega}^a_\mu)\omega^a_\mu$ (with ghost number $(-1)+1$) 
are  the fermionic (anti-)ghost fields corresponding to tensor field $B^a_{\mu\nu}$.  
The bosonic scalar fields $(\bar \beta^a) \beta^a$
(with ghost number $(-2)+2$) are the (anti-)ghost fields for the fermionic vector (anti-)ghost fields 
and $n^a$ is a bosonic scalar ghost field (with ghost number zero). The latter scalar ghost field is
required for the stage-one reducibility of the tensor field. Furthermore,  
$\alpha^a$ and $\bar{\alpha}^a$ are the additional Grassmann valued auxiliary fields (having ghost 
number $+1$ and $-1$).  This model contains a massive non-Abelian gauge field and it was shown to be 
BRST invariant~\cite{al:1997, rohit1, rohit2}. In~\cite{rohit1, rohit2}, it is seen that the model 
is also invariant under the anti-BRST symmetry transformations. It is to be noted that the $\mathcal{CP}$ symmetry is not violated in this model.

\section{Vertex rules and propagators of fields}
The propagators for the $A$ and $B$ fields are found from the Abelian $B\wedge F$ model. The Lagrangian density for the Abelian model is  
\begin{eqnarray}
\mathscr{L} = -\frac{1}{4}\,F_{\mu\nu} F^{\mu\nu} + \frac{1}{12}\, H_{\mu\nu\lambda} H^{\mu\nu\lambda} 
+ \frac{m}{4}\,\varepsilon^{\mu\nu\rho\lambda}\, B_{\mu\nu} \,F_{\rho\lambda},
\label{topab}
\end{eqnarray} 
 where $F_{\mu\nu} = \partial_\mu A_\nu - \partial_\nu A_\mu$ is the field strength of the Abelian gauge field $A_\mu$, 
$H_{\mu\nu\lambda} = \partial_\mu B_{\nu\lambda} + \partial_\nu B_{\lambda\mu} + \partial_\lambda B_{\mu\nu}$ is the field strength
for the tensor field $B_{\mu\nu}$ and  $m$ is the coupling constant of the topological term which has dimension of mass 
(in natural units $\hbar = c = 1$).  The Lagrangian density is invariant under the following two independent gauge transformations, namely; 
\begin{eqnarray}
A_\mu \to A_\mu+\partial_\mu \Lambda, \qquad B_{\mu\nu}\to B_{\mu\nu},
\end{eqnarray} 
\begin{eqnarray}
A_\mu\to A_\mu, \qquad B_{\mu\nu}\to B_{\mu\nu}+\partial_{[\mu}\Lambda_{\nu]},
\label{transformations}
\end{eqnarray}  
where $\Lambda(x)$ and $\Lambda_\mu(x)$ are scalar and vector gauge transformation parameters which vanishes at infinity. 
The Euler-Lagrange equations of motion derived from the above Lagrangian density are as follows
\begin{eqnarray} 
\partial_\mu F^{\mu\nu} &=& - \frac{m}{6} \varepsilon^{\nu\mu\lambda\kappa} \, H_{\mu\lambda\kappa},\nonumber\\ 
\partial_\mu H^{\mu\nu\lambda} &=& + \frac{m}{2}\, \varepsilon^{\nu\lambda\kappa\rho}\, F_{\kappa\rho}.
\end{eqnarray} 
It is interesting to note that one can decouple the above equations for the gauge fields in the following way
\begin{eqnarray}
\big(\Box + m^2 \big)F_{\mu\nu} = 0, \qquad  \big(\Box + m^2 \big)H_{\mu\nu\lambda} =0, 
\end{eqnarray}  
which shows the well-known Klein-Gordon equations for the massive fields $A_\mu$ and $B_{\mu\nu}$.

\begin{figure}[h!]
\begin{center}
\includegraphics[scale=0.04]{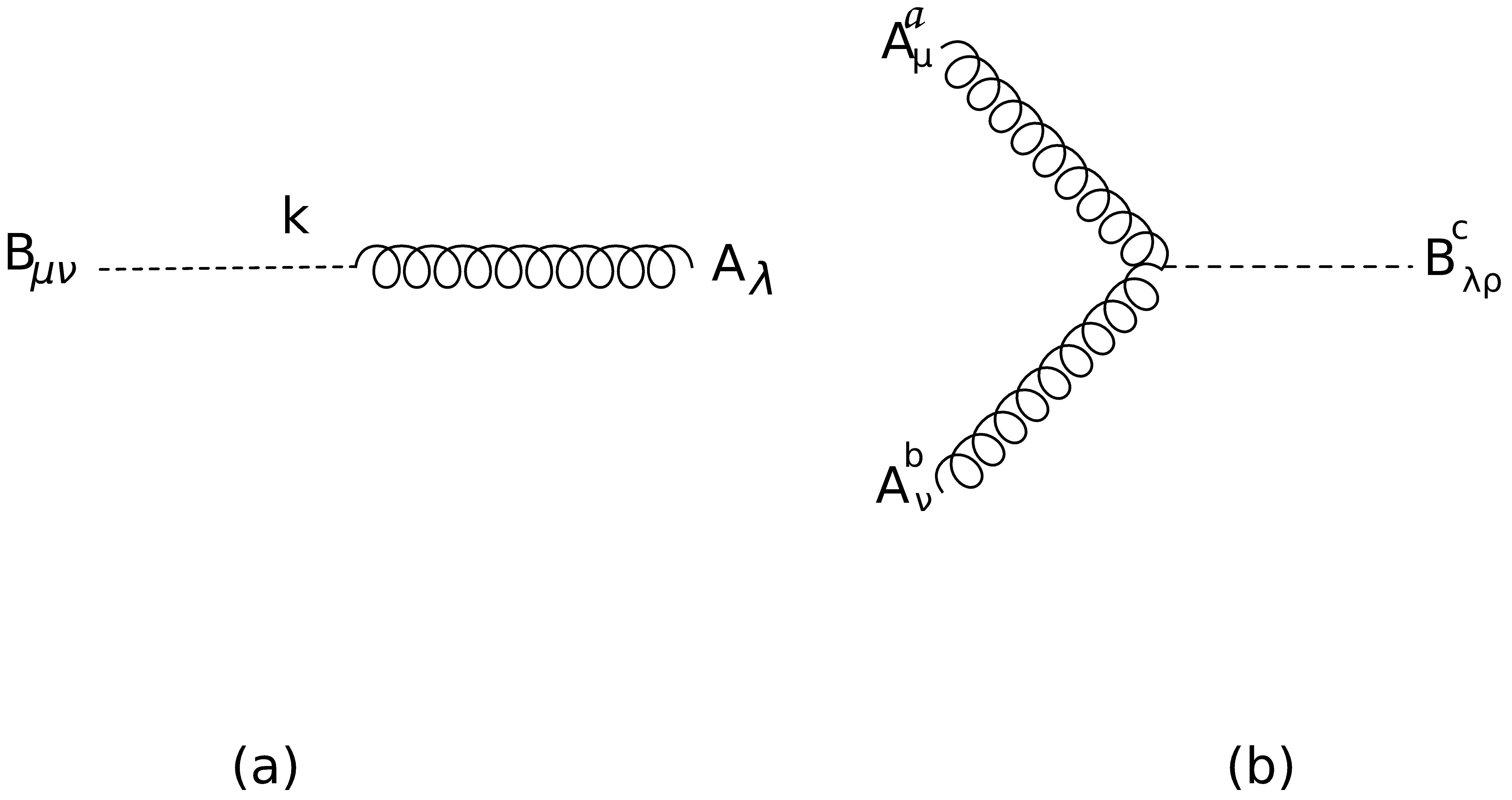} 
\caption{ (a) $B-A$ vertex and (b) $ABB$ vertex from $B\wedge F$ term}
\label{fig:BA-vertex}
 \end{center} 
\end{figure}

\begin{figure}[h!]
\begin{center}
    \includegraphics[scale=0.4]{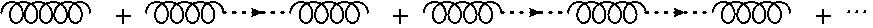} 
\caption{Massive $A$ propagator by summing over $B$ insertions}
  \label{fig:Wprop}
  \end{center}
\end{figure}

\begin{figure}[h!]
  \begin{center}
    \includegraphics[scale=0.2]{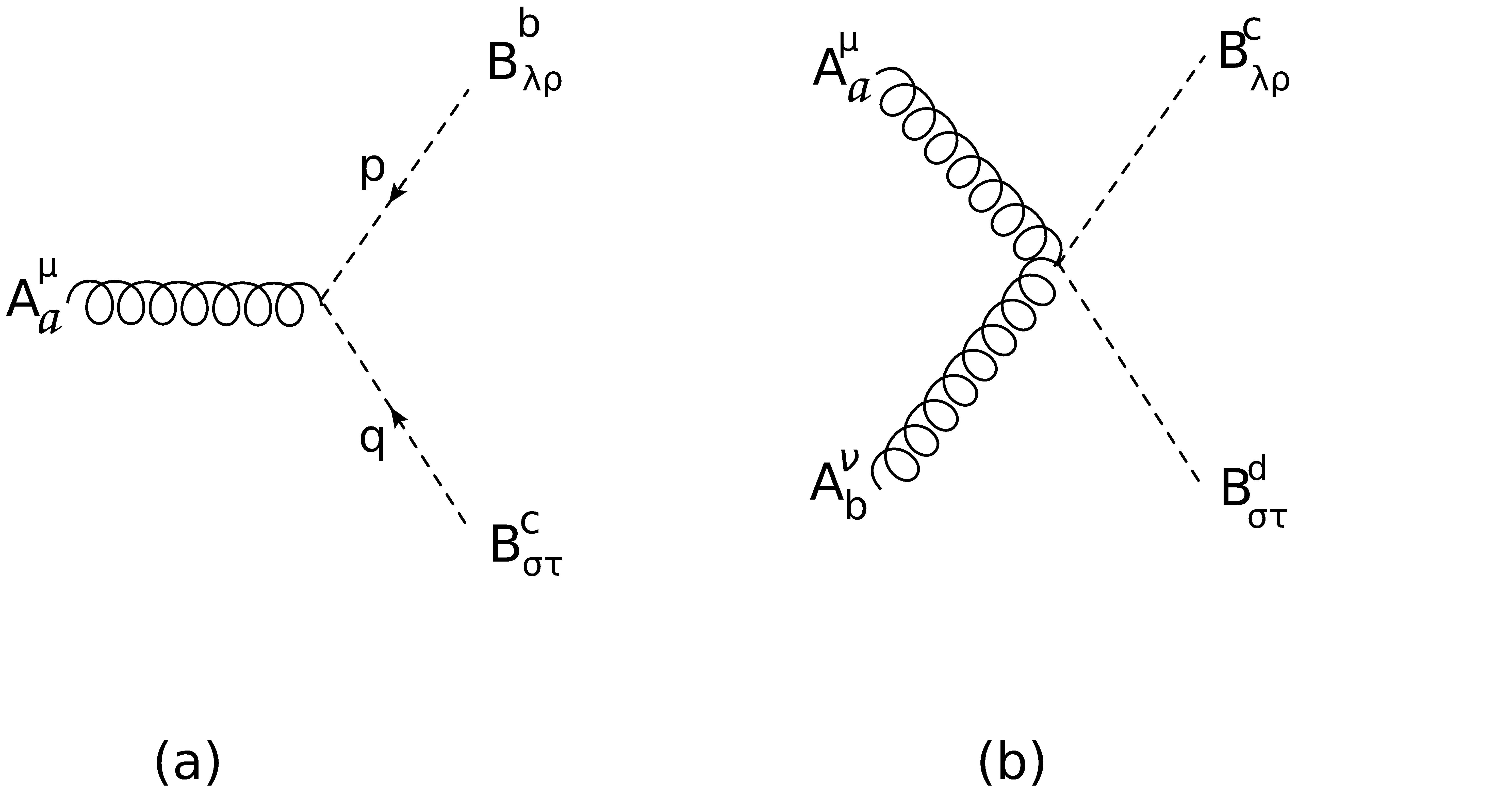} 
\caption{$ABB$ and $AABB$ vertices}
  \label{t4vertices}
  \end{center}
\end{figure}

\begin{figure}[h!]
\begin{center}
\includegraphics[scale=0.04=3]{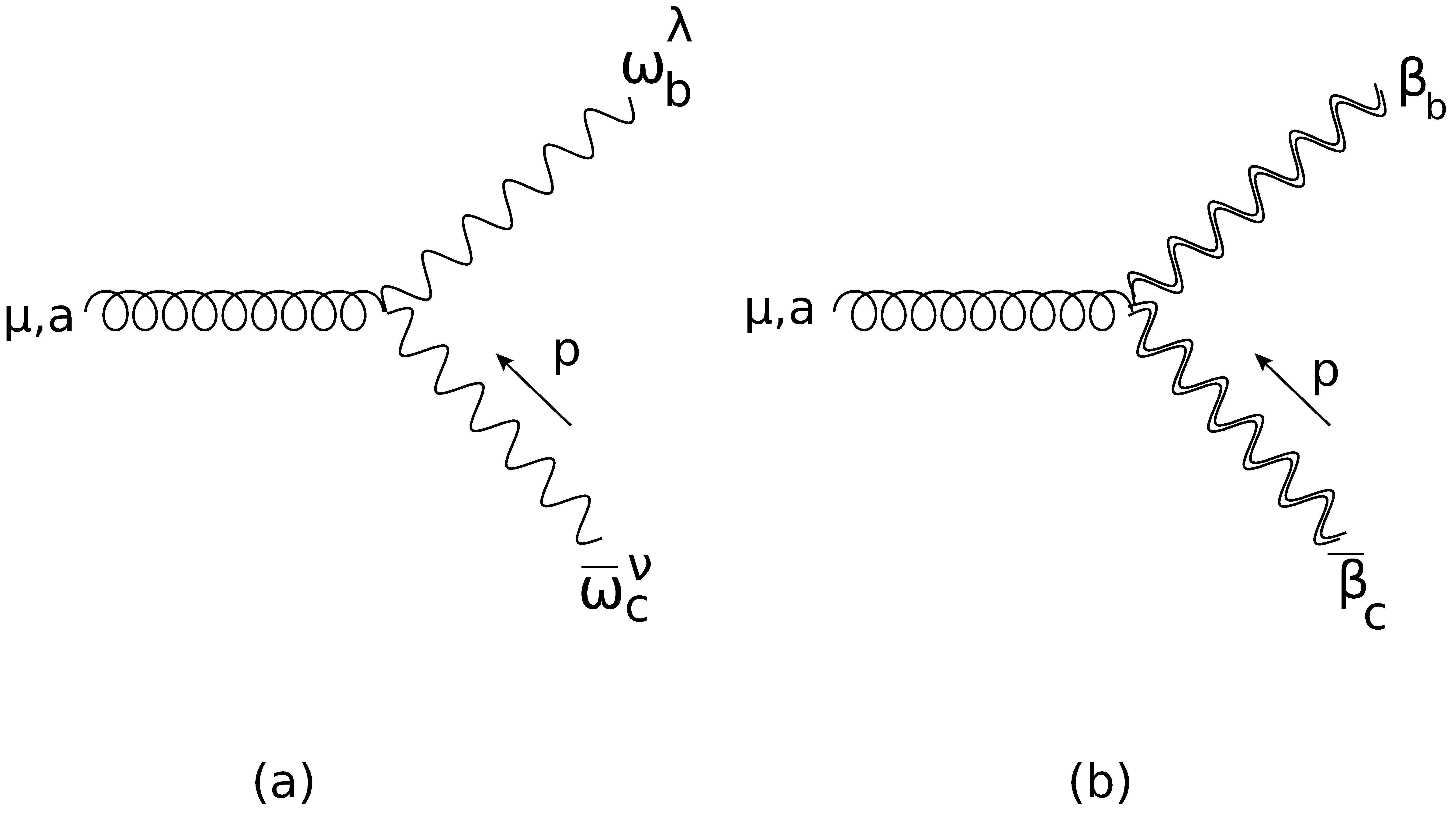} 
\caption{(a) Trilinear vertex among $A_\mu$, $\bar{\omega}_\nu$ and $\omega_\lambda$; (b) trilinear vertex 
among $A_\mu$, $\beta$ and $\bar{\beta}$. Wavy line designates the vector ghosts in (a) and ghosts of 
the vector ghost fields are represented as double wavy line in (b)}
\label{avecg}
\end{center}
\end{figure}

We will consider loop calculation which requires the propagators of $A_\mu$ and $B_{\mu\nu}$ fields. To achieve this,  we introduce the gauge-fixing terms in the  Lagrangian density given
in Eq.~(\ref{topab}) as
\begin{eqnarray}
\mathscr{L}_{gf} = - \frac{1}{2\xi}(\partial_\mu A^\mu)^2 + \frac{1}{2\eta}(\partial_\mu B^{\mu\nu})^2,
 \end{eqnarray}
where $\xi$ and $\eta$ are the gauge-fixing parameters. The topological term is also  quadratic in nature containing both
$A_\mu$ and $B_{\mu\nu}$ fields. To calculate the propagator of the fields, we should take all the quadratic 
terms in the Lagrangian density, excluding  $B\wedge F$ term. The propagators of $A_\mu$ and $B_{\rho\lambda}$ fields are given by
\begin{eqnarray}
i\Delta_{\mu\nu} &=& -\frac{i}{k^2}\left(\eta^{\mu\nu} - (1 - \xi)\frac{k^\mu k^\nu}{k^2}\right), \\
i\Delta_{\mu\nu,\rho\lambda} &=& \frac{i}{k^2}\left(\eta_{\mu [\rho}\, \eta_{\lambda] \nu} 
- (1 - \eta)\,\frac{k_{\mu} \, k_{[\lambda} \,\eta_{\rho]\nu} - k_{\nu} \,k_{[\lambda}\, \eta_{\rho]\mu}} {k^2}\right) .
\label{prop}
\end{eqnarray}
The vertex for the interaction term containing these fields is given by 
%
%
\begin{eqnarray}
iV_{\mu\nu, \lambda} &=& -m\, \varepsilon_{\mu\nu\lambda\rho}\,k^\rho, 
\label{BAvertex}
\end{eqnarray}
which is shown in Fig.~\ref{fig:BA-vertex}.
The complete propagator for the vector field $A_\mu$ can be obtained by taking an infinite number of insertions of the $BA$-vertex 
and the $B$ propagator (cf.  Eq.~(\ref{prop}). This  process is shown in the Fig.~\ref{fig:Wprop} and the sum of 
diagrams can be written as the infinite sum as shown in Fig.~\ref{fig:Wprop}.
 Thus, the complete propagator for  massive vector bosons is given by 
\begin{eqnarray}
iD_{\mu\nu} & = & i \Delta_{\mu\nu} + i \Delta_{\mu\mu'}\,\frac{1}{2}\, iV^{\sigma\rho, \mu'}\, i
\Delta_{\sigma\rho, \sigma'\rho'} \,
\frac{1}{2}\, iV^{\sigma'\rho', \nu'}\, i \Delta_{\nu'\nu} + \cdots \nonumber \\ 
&=& - i \left[\frac{\eta_{\mu\nu} - (1 - \xi)\, \frac{k_\mu \,k_\nu}{k^2}}{(k^2 - m^2)} - \xi\, m^2 \,\frac{k_\mu \,k_\nu}{k^4 (k^2 - m^2)}\right],
\label{aprop}
\end{eqnarray}
where $m$ clearly represents the mass of vector gauge bosons. 
The factors of $\frac{1}{2}$  compensate for double-counting due to the anti-symmetrization of the indices. 
Similarly, for the tensor field $B$, we have the following propagator
\begin{eqnarray}
iD_{\mu\nu, \rho\lambda} = \left[\frac{\eta_{\mu[\rho} \, \eta_{\lambda]\nu} 
+ (1 - \eta)\,\frac{k_{[\mu} \,k_{[\lambda}\, \eta_{\rho]\nu]}}{k^2}}{k^2 - m^2}
+ \eta \,m^2 \,\frac{k_{[\mu}\, k_{[\lambda}\, \eta_{\rho]\nu]}}{k^4 (k^2 - m^2)}\right].
\label{bprop}
\end{eqnarray}
The kinetic term of the YM field (cf. Eq.~(\ref{nonabactn})) provides the derivative trilinear and  quartic couplings. 
The interaction part of the kinetic term of YM field  is 
\begin{eqnarray}
\mathcal{L}_{int} = \frac{1}{4}\,g f^{bca}\, A^{\mu b}\, A^{\nu c} \, \left(\partial_{[\mu} \,A_{\nu]}^a 
- 2 g f^{dea}\, A^d_\mu \,A^e_\nu\right).
\end{eqnarray}
 \begin{figure}[h!]
\begin{center}
    \includegraphics[scale=0.04]{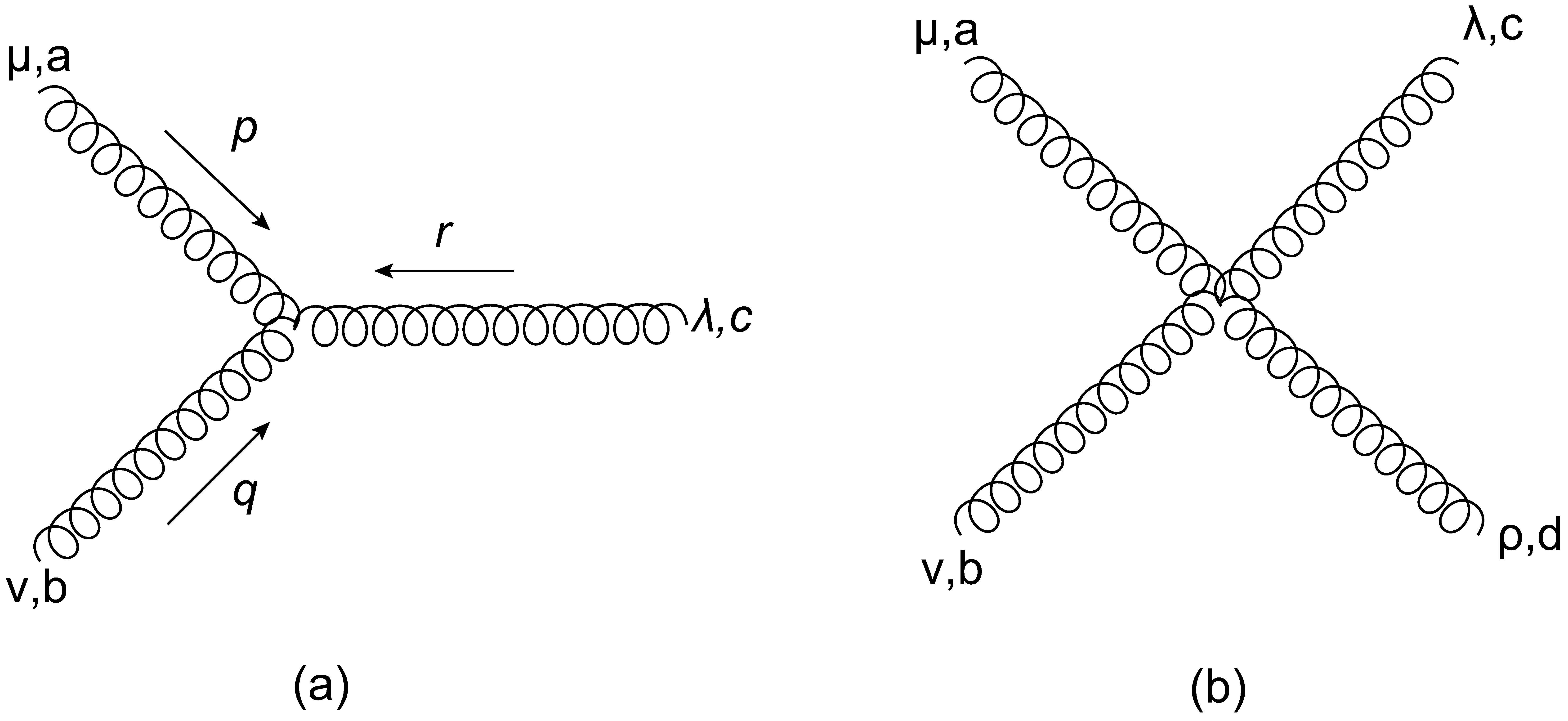} 
\caption{(a) $AAA$- trilinear vertex; (b) $AAAA$ quartic vertex}
  \label{ymvertices}
  \end{center}
\end{figure}
The vertex rules corresponding to these couplings are as follows
\begin{eqnarray}
V^{abc}_{\mu\nu\lambda} &=& -g f^{abc}\left[\left(q - r\right)_\mu\, \eta_{\nu\lambda} + \left(r - p\right)_{\nu}\, \eta_{\lambda\mu}
+ (p - q)_{\lambda}\, \eta_{\mu\nu}\right],\\
V^{abcd}_{\mu\nu\lambda\rho}&=&-ig^2\left[f^{abe}\,f^{cde}\, \eta_{\mu[\lambda} \, \eta_{\rho]\nu}
+ f^{ace}\, f^{bde}\, \eta_{\mu[\nu} \, \eta_{\rho]\lambda} + f^{ade}\,f^{bce}\, \eta_{\mu[\nu}\, \eta_{\lambda]\rho}\right],
\end{eqnarray}
where $f$'s are the structure constants of $SU(N)$ group, which are totally antisymmetric in their indices. The momenta of the particles at the trilinear 
vertex is shown in Fig.{ymvertices}a. The topological term also provides a trilinear coupling $ABB$ with vertex term
%
%
\begin{eqnarray}
iV_{\mu,\nu, \lambda\rho}^{abc} &=& -ig \,m \,f^{bca}\, \varepsilon_{\mu\nu\lambda\rho}\,, 
\end{eqnarray}
To proceed further,  we require 
the propagators of vector ghost fields, $\omega^\mu$,  $\bar{\omega}^\mu$,  the ghost fields of the vector 
ghost fields, $\beta$, $\bar{\beta}$, and the ghost fields corresponding to the one form gauge field, $\omega$, 
$\bar{\omega}$ appeared in Eq.~(\ref{nonabactn}). We can get the vertex factors for trilinear and quartic couplings $ABB$ and $AABB$ as
\begin{eqnarray}
iV^{abc}_{\mu, \lambda\rho, \sigma\tau} &=& g \,f^{abc}\left[(p - q)_\mu\,\eta_{\lambda  [\sigma} \, \eta_{\tau]\rho} 
+ p_{[\sigma} \, \eta_{\tau][\lambda}\, \eta_{\rho]\mu} - q_{[\lambda} \, \eta_{\rho][\sigma}\, \eta_{\tau]\mu}  \right], \\
iV^{abcd}_{\mu, \nu, \lambda\rho, \sigma\tau} &=& ig^2 \bigg[f^{ace}\, f^{bde} 
\left(\eta_{\mu\nu}\, \eta_{\lambda\, [\sigma}\, \eta_{\tau]\rho} + \eta_{\mu\, [\sigma}\, \eta_{\tau][\lambda}\, \eta_{\rho]\nu} \right) \nonumber\\
&+& f^{ade}\, f^{bce}\left(\eta_{\mu\nu}\, \eta_{\lambda\, [\sigma}\, \eta_{\tau]\rho} + \eta_{\mu\, [\lambda}\, \eta_{\rho][\sigma} \, \eta_{\tau]\nu}\right) \bigg],
\end{eqnarray}
where the vertices are shown in Fig.~\ref{t4vertices}. The propagator of vector ghost
 field from the Lagrangian density 
\begin{eqnarray}
\mathcal{L} = -\partial_\mu \bar{\omega}_\nu^a\left(\partial^\mu\omega^{\nu a} - \partial^\nu \omega^{\mu a}\right)
-\frac{1} {\tilde{\xi}}\, \left(\partial_\mu \omega^{\mu a}\right)\left(\partial_\nu \bar{\omega}^{\nu a}\right).
\end{eqnarray}
at the gauge $\tilde{\xi}=1$ can be obtained \textcolor{red}{by} integrating out $\alpha$ and $\bar{\alpha}$ from the action in Eq.~(\ref{nonabactn}). 
\begin{figure}[h!]
        \centering
        \begin{minipage}[c]{.25\textwidth}
             \includegraphics[width=\linewidth]{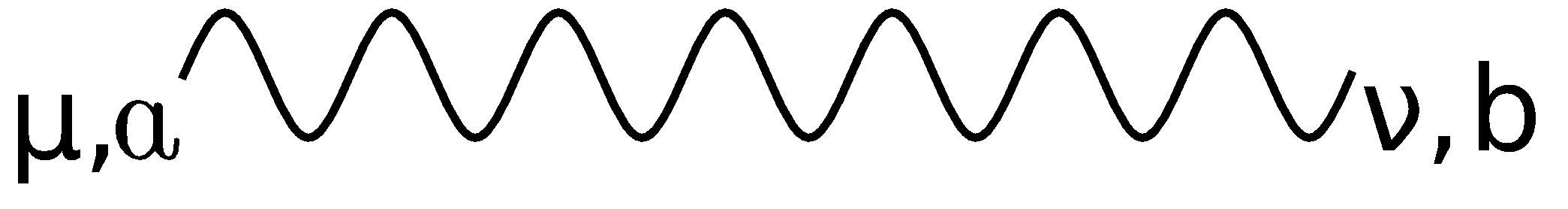}
            \label{fig:ghostprop}
        \end{minipage}%
        \hspace{0.8cm}
        \begin{minipage}[c]{.15\textwidth}
          \begin{equation}
            \nonumber{i\Delta_{\omega\bar{\omega}, ab}} = - \frac{i}{p^2} \left[\eta^{\mu\nu}-(1-\tilde{\xi}) \frac{p^\mu p^\nu}{p^2}\right] \delta_{ab}
            \label{eq:ghostprop}
            \end{equation}
            \end{minipage}
         \end{figure}
\begin{figure}[h!]
       \centering
        \begin{minipage}[c]{.25\textwidth}
           \includegraphics[width=\linewidth]{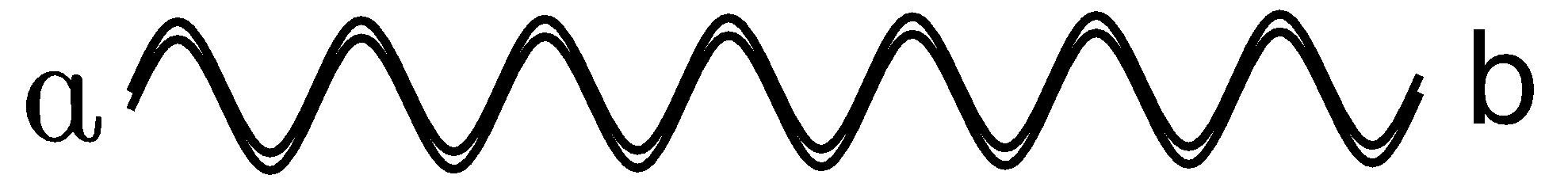}
            \label{fig:betaprop}
        \end{minipage}%
       \hspace{2.0cm}
        \begin{minipage}[c]{.15\textwidth}
          \begin{equation}
          \nonumber  i\Delta_{\bar{\beta},\beta, ab}= \frac{i}{p^2}\delta_{ab}
            \label{eq:betaprop}
            \end{equation}
            \end{minipage}
    \end{figure} 
Apart from the usual trilinear coupling among Fadeev-Popov ghost and YM fields, we can also see the 
action contains another trilinear coupling among vector ghosts and YM fields which is given by
\begin{eqnarray}
\mathcal{L}^{vec-gh-A}_{int} = - g\,f^{abc}\,\partial^{\nu} \bar{\omega}^{\mu a} A^b_{[\mu}\, \omega_{\nu]}^c.
\label{Avecghcoupling}
\end{eqnarray}
The vertex factor corresponding to Lagrangian density given in Eq.~(\ref{Avecghcoupling}) is
\begin{eqnarray}
i V^{abc}_{\mu\nu\lambda} = -g\, f^{abc}\left(p_\nu\, \eta_{\mu\lambda} - p_\mu \, \eta_{\nu\lambda}\right).
\end{eqnarray}
In derivation of the above rule, all the four momentums are taken as incoming towards the vertex as shown 
in Fig.~\ref{avecg}(a). There is also a trilinear coupling
among YM and ghost of the vector ghost fields. The trilinear vertex is shown in Fig.~\ref{avecg}b.  The coupling 
is given by the following Lagrangian density as 
\begin{eqnarray}
\mathcal{L}^{A\beta\bar{\beta}}_{int} = - g\, f^{bca} A_\mu^b\, \partial^\mu \, \bar{\beta}^a \, \beta^c,
\label{Abb}
\end{eqnarray}
Since, the coupling in Eq.~(\ref{Abb}) also  contains derivative of fields, the  corresponding vertex  will be momentum dependent. 
 This trilinear coupling is same as the trilinear coupling among YM field and its 
FP ghosts with the vertex term and it is given by
\begin{eqnarray}
iV^{\mu}_{abc} =- g f_{abc} p^\mu.
\end{eqnarray}


\section{One loop correction}
 
Using the vertices and  the  propagators, we calculate the one loop correction of the soft modes of 
massive gluons. These calculations are done in the Feynman-`t Hooft gauge: $\xi = \eta = \tilde{\xi} = 1$ in $4D$ Euclidean space, 
where  Minkowskian metric $\eta^{\mu\nu}$ is replaced by Euclidean metric $\delta^{\mu\nu}$.
We consider a generic form of the loop amplitude for the calculation
\begin{eqnarray}
\Pi_{\mu\nu}^{ab} = \frac{g^2 N_C}{n}\, \delta^{ab}\,\sumint_\textbf{p} \frac{\delta_{\mu\nu} \left(a_1\, p^2 + a_2\, k^2 + a_3\,  m^2\right)
+ a_4\, k_\mu \,k_\nu + a_4 (p_\mu \, k_\nu + p_\nu \, k_\mu) + a_5\, p_{\mu} \,p_{\nu}}{\left(p^2 - m^2\right)
\left\{\left(k - p\right)^2 - m^2\right\}},
\label{genint}
\end{eqnarray}
where $\displaystyle\sumint_\textbf{p}\equiv \displaystyle\sumint_{\vec{p}}= \displaystyle\sum_{p_{0n}}T\int_\textbf{p}, ~\int_\textbf{p}=\int\frac{d^d\textbf{p}}{(2\pi)^d}$. 
We assume that the external legs carry soft momenta$\sim gT$. We take the Matsubara sum over the temporal component $p_{on}$ of the four momentum $p_{\mu}$ and integrate over spatial component $\textbf{p}\equiv\{p_i\}$. To carry
out further, we approximate the energy of the external legs $E_1$ and $E_2$ as follows
\begin{eqnarray}
E_1 &=& \sqrt{P^2 + m^2} \approx P + \frac{m^2}{2P} + \cdots, \label{expan1}\\
E_2 &=&\sqrt{\left(P - K\right)^2 + m^2} \approx P - \textbf{k} \cdot \textbf{v} + \frac{m^2}{2P} \label{expan2},
\end{eqnarray}
where $|\vec{p}|=P$ and $v_i=\dfrac{p_i}{P}$. Now we consider
\begin{eqnarray}
\mathcal{G} &=& T \, \sum_{p_{0n}}\, \frac{1}{\left(p_{0n}^2 + E_1^2\right)\left\{(k - p)_{0n}^2 - m^2\right\}},\nonumber\\
\end{eqnarray}
which, summing over $p_{0n}$, results in
\begin{eqnarray}
\mathcal{G}&=&\frac{1}{4E_1 E_2} \Bigg[\frac{1}{ik_{on}-E_1-E_2}\left(-n_B(E_1)-n_B(E_2)-1\right) \nonumber\\
&+& \frac{1}{ik_{on}+E_2-E_1}\left(n_B(E_1)-n_B(E_2)\right) \nonumber\\ 
&+& \frac{1}{ik_{on}+E_1-E_2}\left(n_B(E_2)-n_B(E_1)\right)\nonumber\\
&+& \frac{1}{ik_{on}+E_2+E_1}\left(1+n_B(E_1)+n_B(E_2)\right)\Bigg].
\label{expan}
\end{eqnarray}
  The quadratic terms $\mathcal{O}(p^2)$ are neglected from the numerator  because of the assumption $p\sim gT (g\ll T)$ 
Using Eqs.~(\ref{expan1}) and (\ref{expan2}), the above expression of $\mathcal{G}$ can be approximated as
\begin{eqnarray}
\tilde{\mathcal{G}} &\approx & \left[- \frac{1}{2P} \left\{-2n_B(P)\right\} 
+ \mathbf{k} \cdot \mathbf{v}\, \frac{n'_B(P)}{ik_{on}-\textbf{k} \cdot \textbf{v}} 
- \mathbf{k} \cdot \mathbf{v} \,\frac{n'_B(P)}{ik_{on}+\textbf{k} \cdot \textbf{v}}+\frac{1}{2P} \left\{2n_B(P)\right\}\right].
\label{expan4}
\end{eqnarray}
In the expression of integrand in Eq.~(\ref{genint}), the denominator contains  identical propagators of 
bosons appearing in loop. Hence, the term $\left(p_\mu k_\nu+p_\nu k_\mu\right)$ will be simplified after renaming the variable 
$p\to k-p$, one half of this term as   
\begin{eqnarray}
p_\mu\, k_\nu + p_\nu \, k_\mu &\to & \frac{1}{2} \left[p_\mu \,k_\nu + p_\nu \, k_\mu + (k - p)_\mu \, k_\nu 
+ (k - p)_\nu \, k_\mu\right] = k_\mu \, k_\nu.
\label{trick}
\end{eqnarray}  
After rearranging the terms in the numerator of the integrand in Eq.~(\ref{genint}) and neglecting the  term 
$\sim\mathcal{O}(K^2)$ then the spatial part reads as  
\begin{eqnarray}
N_{ij}^{ab} &=& \frac{g^2 N_C \, \delta^{ab}}{n}\left[\delta_{ij}\left(a_1 p^2 + a_3 m^2\right) + a_5 p_{i}p_{j}\right] \nonumber\\ 
&=& \frac{g^2 N_C\, \delta^{ab}}{n}\left[\delta_{ij}\left(a_1( p^2-m^2)+a_1m^2 +a_3 m^2\right)+a_5 p_{i}p_{j}\right] \nonumber \\
&=& \frac{g^2 N_C \, \delta^{ab}}{n}\left[\delta_{ij}\left(a_1( p^2-m^2)+(a_1+a_3)m^2 \right)+a_5 p_{i}p_{j}\right]. 
\label{pi1}
\end{eqnarray}
From the above expression, it is clear that the presence of the term $m^2\delta_{ij}$ will provide  magnetic 
mass of gluons. In constructing an effective field theory, we neglect the quadratic term of $\mathcal{O}(K^2)$ from the above 
expression. Substituting $\mathcal{G}$ in Eq.~(\ref{expan4}) the spatial part of the $\Pi_{\mu\nu}$ which, after  
changing the variable $\textbf{v} \to - \textbf{v}$, reads,
\begin{eqnarray}
\Pi_{ij}^{ab} \approx  \frac{g^2 N_C}{n} \delta^{ab} \int_\textbf{p}\Bigg[a_1\, \frac{n_B(P)}{P}\, \delta_{ij}
+ \left(\widetilde{A}\, m^2 \delta_{ij} + a_5 \,p_i \, p_j\right)
\left\{\frac{n_B(P)}{2P^3} + \textbf{k} \cdot \textbf{v}\frac{n'_B(P)}{2P^2\left(ik_{0n} 
- \textbf{k} \cdot \textbf{v}\right)}\right\}\Bigg],
\end{eqnarray}
where $\widetilde{A}=(a_1+a_3)$, $\int_p = c(d) \int p^{d - 1} dp$ and $c(d) =\dfrac{2}{(4\pi)^{d/2}\Gamma(d/2)}$. 
The angular integration goes over directions of $v_i\equiv\dfrac{p_i}{P}$ and normalized to unity as:
\begin{eqnarray}
\int d\Omega_v = 1,
\end{eqnarray} 
and using the rotational invariance, we get
\begin{eqnarray}
\int d\Omega_v \, v_i \,v_j = \frac{1}{d}\, \delta_{ij}.
\label{vvint}
\end{eqnarray}
Thus we  are going to construct the effective field theory in the energy scale, $E = \sqrt{P^2 + m^2}$, 
where  $m\ll E\leqslant T$. Using Eq.~(\ref{vvint}) and the following identity for $d=3$
\begin{eqnarray}
\int_P n'_B(P) = - (d - 1)\int_P \frac{1}{P} \,n_B(P),
\end{eqnarray}
and Eq.~(\ref{vvint}), the $\Pi^{ab}_{\mu\nu}$ can be expressed as
\begin{eqnarray}
\Pi _{ij}^{ab} &=& \delta^{ab}\, \frac{g^2 N_c}{n} \Bigg[\int_p \frac{n_B(P)}{P}\Bigg\{\left(a_1 
+ \frac{a_5}{2d}- a_5 \, \frac{(d-1)}{2d}\right)\delta_{ij} 
- \frac{(d - 1)}{2} a_5 \int d\Omega_v \, \frac{v_i v_jk_{0n}}{ik_{0n} - \mathbf{k} \cdot \mathbf{v}}\Bigg\}\nonumber \\
&& \hskip 1cm  +A \, \delta_{ij} \, \int_\mathbf{p} \,\mathbf{k} \cdot \mathbf{v}\, \frac{m^2\, n'_B(P)}{P^2\left(ik_{0n} 
- \mathbf{k} \cdot \mathbf{v}\right)}\Bigg].
\label{Piij}
\end{eqnarray}
The last term of the above integrand can be rearranged  as
\begin{eqnarray}
\int_\mathbf{p}\mathbf{k} \cdot \mathbf{v}\frac{m^2 \,n'_B(P)}{P^2\left(ik_{0n} - \mathbf{k} \cdot \mathbf{v}\right)}
= m^2 I \left(- 1 + k^0\, L(K)\right).
\end{eqnarray}
where
\begin{eqnarray}
L(K) = \int d \Omega_v \,\frac{1}{ik_{0n} - \mathbf{k}\cdot \mathbf{v}},
\label{L(K)}
\end{eqnarray}
and
\begin{eqnarray}
I =-\beta\int_{0}^{\infty}\frac{e^{\beta P}}{(e^{\beta P}-1)^2}\, dP,
\end{eqnarray}
which, after integrating over $P$, in the above,  yields
\begin{eqnarray}
I=\frac{1}{e^{\beta P}-1}\Bigg |_{P=0}.
\label{divint}
\end{eqnarray}
Here the divergence appearing in $I$ is an artifact of the approximations made\footnote{See the detail in appendix A. } in Eq.~(\ref{expan1}) and Eq.~(\ref{expan2}).  

With the help of Eq.~(\ref{expan2}), we can now re-express Eq.~(\ref{Piij}) as
\begin{eqnarray}
\Pi_{ij} &=& \frac{g^2N_C}{n}\bigg[\left\{\widetilde{B} \frac{T^2}{12} + A \, m^2 I
\left(-1+k^0 L(K)\right)\right\}\delta_{ij} \nonumber\\
&& \hskip 1cm + \,C\, \frac{ T^2}{12}\left(\mathcal{P}^T_{ij} \, \Pi_T(K)+\mathcal{P}^E_{ij} \, \Pi_E(k)\right)\bigg],
\label{genform}
\end{eqnarray}   
where in $d$ spatial dimensions,  $\widetilde{B} = \left(a_1 + \dfrac{a_5}{2d} - \dfrac{a_5(d-1)}{2d} \right)$ and $C = - \dfrac{(d - 1)}{2} a_5$. 
The factor $\dfrac{T^2}{12}$ appears from the integration $\displaystyle\int_P \dfrac{1}{P}n_B(P)$, where 
$\displaystyle\int_P\equiv V(d)\int_{0}^\infty p^{d-1} dp$ and $V(d)=\dfrac{2}{(4\pi)^{\frac{d}{2}}\Gamma\left(\frac{d}{2}\right)}$. 
The gauge indices in the above calculations has been suppressed and the coefficients of the projection operators are found as
\begin{eqnarray}
\mathcal{P}^T_{\mu\nu}(k) &\equiv & \delta_{\mu i}\, \delta_{\nu j} P^T_{ij}(k),\\ 
\mathcal{P}^E_{\mu\nu} &\equiv & \delta_{\mu\nu} - \frac{k_\mu \, k_\nu}{k^2} - \mathcal{P}^T_{\mu\nu}(k),
\end{eqnarray} 
where $P^T_{ij}(k)=\delta_{ij} - k_i\, k_j/K^2$ and in the 3-dimension~\cite{lain}
\begin{eqnarray}
\Pi_T(K) &=& \frac{1}{2}\left[\frac{(ik_{0n})^2}{K^2} + \frac{ik_{0n}}{2K}
\left\{1 - \frac{(ik_{0n})^2}{K^2}\right\}\ln\frac{ik_{0n} + K}{ik_{0n} - K}\right], \\
\Pi_E(K) &=& \left[1 - \frac{(ik_{0n})^2}{K^2}\right]\left[1 - \frac{ik_{0n}}{2K}\ln\frac{ik_{0n} + K}{ik_{0n} - K}\right].
\end{eqnarray}
Now we can write the effective Lagrangian density as
\begin{eqnarray}
\mathcal{L}_{eff} &=& -\frac{1}{4}F^{\mu\nu}_a F_{\mu\nu}^a+\int_K \tilde{m}^2(K) A^\mu(K) A_\mu (-K) \nonumber\\
&& \hskip 1cm +\, m_E^2\int d\Omega_v \left(\frac{1}{\mathcal{V} \cdot D}\mathcal{V}^\alpha F^a_{\alpha\mu}\right)
\left(\frac{1}{\mathcal{V} \cdot D}\mathcal{V}^\beta F_{a\beta}^\mu\right),
\end{eqnarray}
where $\tilde{m}^2= \dfrac{g^2N_c}{n}\left[B \dfrac{T^2}{12}+  A m^2 I \left(-1+k^0 L(K)\right)\right]$,  $m_E^2 \approx g^2 C \dfrac{N_c}{n} \dfrac{T^2}{12}$ and 
$\mathcal{V}^\alpha\equiv \left(1, \mathbf{v}\right)$.  We have obtained a generic form  of the Debye mass and observed how 
 the ``bare'' mass of gluon contributes in the construction of effective action. Now we are going to consider the relevant contributions to the effective 
Lagrangian from various loop diagrams of the topologically massive model.  The generic form of the 
loop integration is given as
\begin{eqnarray}
\Pi_{ij} &=& m^2 \sumint_{\mathbf{k}} \frac{\left(A \delta_{ij}\, k^2 + B k_i \,k_j\right)}{k^2\left\{(p - k)^2 - m^2\right\}\left(k^2 - m^2\right)},
\end{eqnarray}
which could be written as 
\begin{eqnarray}
\Pi_{ij} &=& m^2 \int_{\mathbf{k}} \left(A \delta_{ij} \, k^2 + B k_i \, k_j\right)\mathcal{G}(E_1, E_2, E_3, k),
\end{eqnarray}
and
\begin{eqnarray}
\mathcal{G}(E_1, E_2, E_3, k) &=& \sum_n \frac{1}{\left(k_{0n}^2 + E_1^2\right)\left\{(p - k)_{0n}^2 + E_2^2\right\}(k_{0n}^2 + E_3^2)}\nonumber\\ 
&=& \frac{1}{E_3^2-E_1^2}\left[\mathcal{G}(E_1, E_2, k)-\mathcal{G}(E_2, E_3, k)\right], \label{3prop}
\end{eqnarray}
where 
\begin{eqnarray}
\mathcal{G}(E_1, E_2, k) &=& T\sum_{p_{0n}}\frac{1}{\left(p_{0n}^2+E_1^2\right)\left\{r_{0n}^2+E_2^2\right\}}\nonumber\\ 
&=& \frac{1}{4E_1 E_2}\Bigg[\frac{1}{ik_{on}-E_1-E_2}\left(-n_B(E_1)-n_B(E_2)-1\right) \nonumber\\ 
&&+ \frac{1}{ik_{on}+E_2-E_1}\left(n_B(E_1)-n_B(E_2)\right) \nonumber\\
&&+ \frac{1}{ik_{on}+E_1-E_2}\left(n_B(E_2)-n_B(E_1)\right) \nonumber\\ 
&&+ \frac{1}{ik_{on}+E_2+E_1}\left(1+n_B(E_1)+n_B(E_2)\right)\Bigg].
\label{expan3}
\end{eqnarray}
Thus, from Eqs.~(\ref{3prop}) and (\ref{expan3}), we get 
\begin{eqnarray}
\mathcal{G}&=&\frac{1}{m^2}\Bigg[\frac{1}{4E_1 E_2}\Bigg\{\frac{1}{ik_{on}-E_1-E_2}\left(-n_B(E_1)-n_B(E_2)-1\right) \nonumber\\ 
&+& \frac{1}{ik_{on}+E_2-E_1}\left(n_B(E_1)-n_B(E_2)\right) + \frac{1}{ik_{on}+E_1-E_2}\left(n_B(E_2)-n_B(E_1)\right)\nonumber\\ 
&+& \frac{1}{ik_{on}+E_2+E_1}\left(1+n_B(E_1)+n_B(E_2)\right)\Bigg\}\nonumber\\ 
&-& \frac{1}{4E_3 E_2}\Bigg\{\frac{1}{ik_{on}-E_2-E_3}\left(-n_B(E_2)-n_B(E_3-\mu)-1\right)\nonumber \\ 
&+& \frac{1}{ik_{on}+E_2-E_3}\left(n_B(E_3)-n_B(E_2-\mu)\right) + \frac{1}{ik_{on}+E_3-E_2}\left(n_B(E_2)-n_B(E_1)\right) \nonumber\\ 
&+& \frac{1}{ik_{on}+E_2+E_3}\left(1+n_B(E_3)+n_B(E_2)\right)\Bigg\}\Bigg],
\end{eqnarray}
where $\dfrac{1}{m^2}$ originates from $E_3^2- E_1^2=k_{0n}^2+m^2-k_{0n}^2=m^2$. Taking the HTL approximation, we can write the above expression in the following form
\begin{eqnarray}
\mathcal{G} \approx \frac{n_B(E_1) - n_B (E_3)}{4K^2 m^2}
\left[\frac{1}{ip_{0n} - \mathbf{p} \cdot \mathbf{v}} - \frac{1}{ip_{0n} + \mathbf{p} \cdot \mathbf{v}}\right].
\label{gfor3prop}
\end{eqnarray}
\begin{figure}[h!]
  \begin{center}
    \includegraphics[scale=0.06=3]{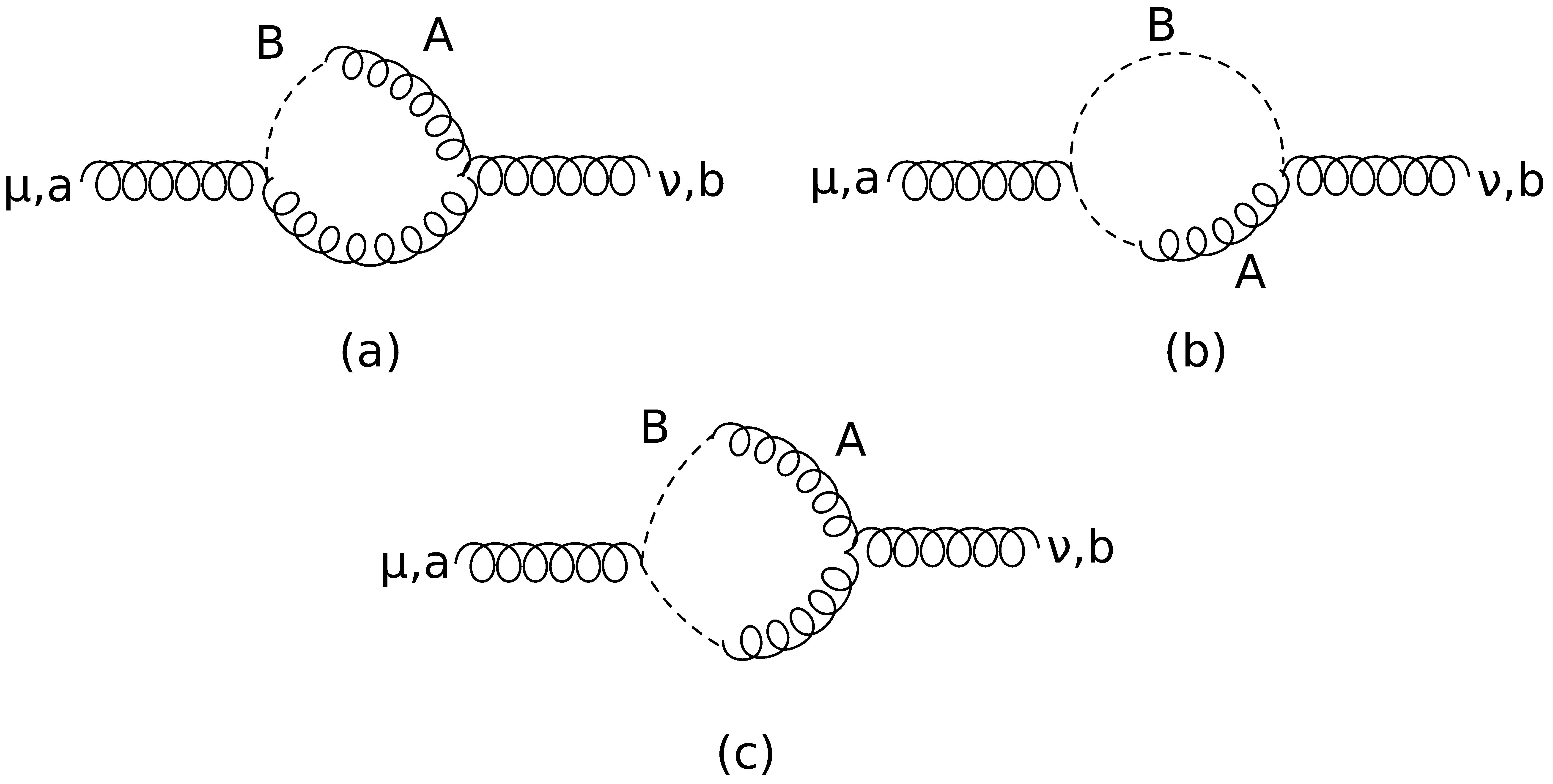} 
\caption{The loops which do not contribute in HTL approximation.}
  \label{loopneg}
  \end{center}
\end{figure}
The  task will becomes simple with the observation  that the diagrams in Fig.~\ref{loopneg} are to be neglected 
in HTL approximation. In this approximation $m\ll K$, then  $n_B(E_1)\approx n_B(E_3)$ at leading order. Hence, the 
contribution to the quantum corrections from the diagrams in Fig.~\ref{loopneg}a and Fig.~\ref{loopneg}b is given by $\Pi_{ij}^{\ref{loopneg}a, 
\ref{loopneg}b}\approx 0$. A similar conclusion can be drawn for the contribution from Fig.~\ref{loopneg}c which contains four propagators. 
Therefore, the $\Pi_{ij}^{\ref{loopneg}c}$ have the Matsubara sum as
\begin{eqnarray}
\mathcal{G}(E_1, E_2, E_3, E_4, k) = \sumint_\mathbf{K}\, 
\frac{1}{\left(k_{0n}^2+E_1^2\right)\left\{(p - k)_{0n}^2 + E_2^2 \right\}(k_{0n}^2 + E_3^2)\left\{(p - k)_{0n}^2 + E_4^2\right\}}.
\end{eqnarray}
After some algebraic computations and HTL approximation, the above expression becomes
\begin{eqnarray}
\mathcal{G}(E_1, E_2, E_3, E_4, k) &\approx& \frac{\left(n_B (E_3) - n_B (E_1)\right)}{4K^2\, m^4}
\Bigg[\left(\frac{1}{ip_n - \mathbf{p} \cdot \mathbf{v}} - \frac{1}{ip_n + \mathbf{p} \cdot \mathbf{v}}\right) \nonumber\\
&+& \left(\frac{1}{ip_n + \mathbf{p} \cdot \mathbf{v}}-\frac{1}{ip_n - \mathbf{p} \cdot \mathbf{v}} \right)\Bigg],
\end{eqnarray} 
which shows that $\Pi_{ij}^{\ref{loopneg}c}$  does not contribute too.

\begin{figure}[h!]
\begin{center}
\includegraphics[scale=0.25]{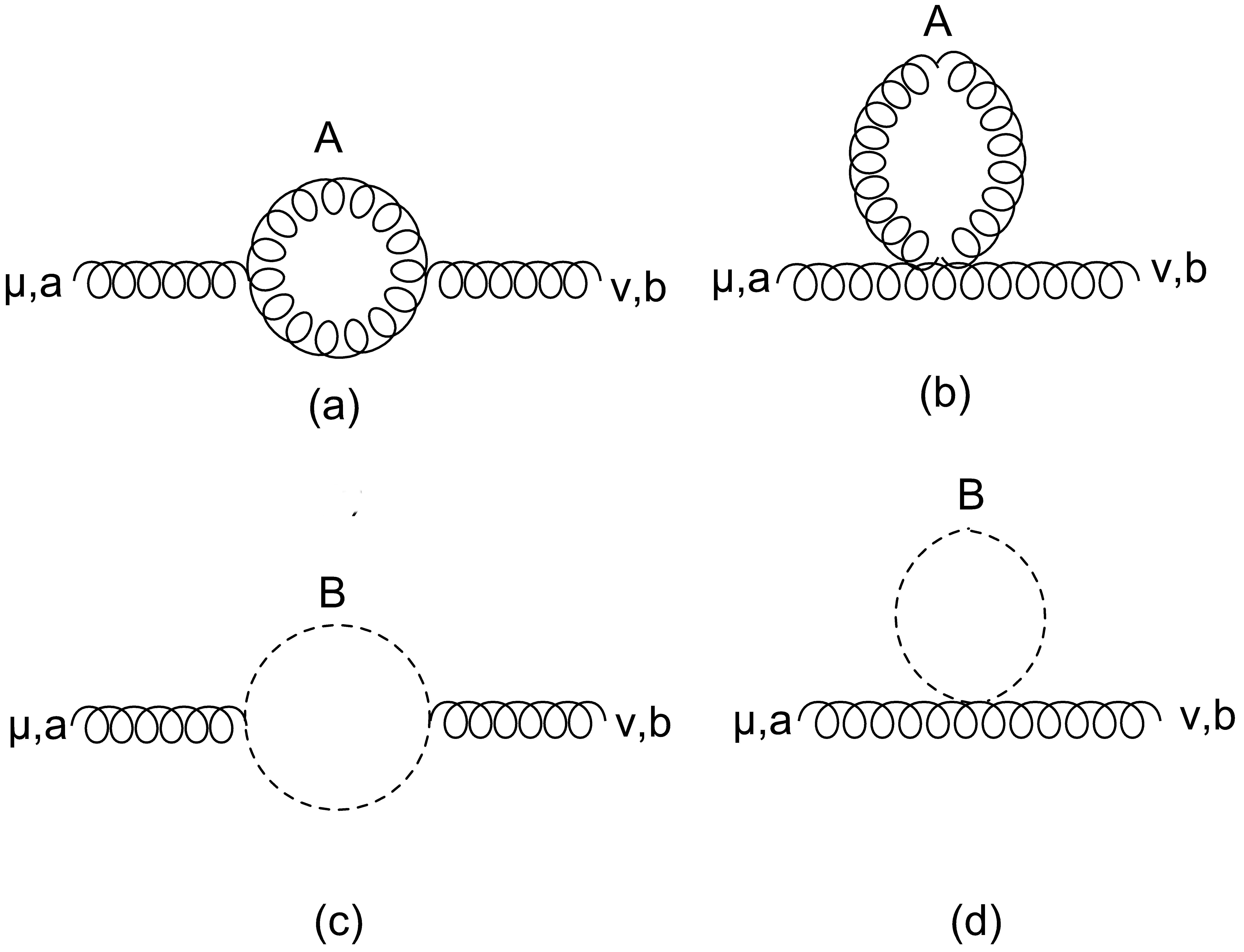} 
\caption{Loop diagrams containing the $A$ and $B$ fields.}
  \label{looprelvnt}
  \end{center}
\end{figure}

  
Only relevant loop diagrams with non-zero contribution, constructed from $A$ and $B$ fields are shown in the Fig.\ref{looprelvnt}.
The rest of the diagrams are from the ghost sectors, where loops are constructed by FP ghost of YM field, 
$\omega$, $\bar{\omega}$; vector ghost $\omega^\mu$, $\bar{\omega}^\mu$ and ghost of the vector ghost $\beta$ and $\bar{\beta}$ corresponding to  tensor field $B_{\mu\nu}$.

\begin{figure}[h!]
\begin{center}
\includegraphics[scale=0.035=3]{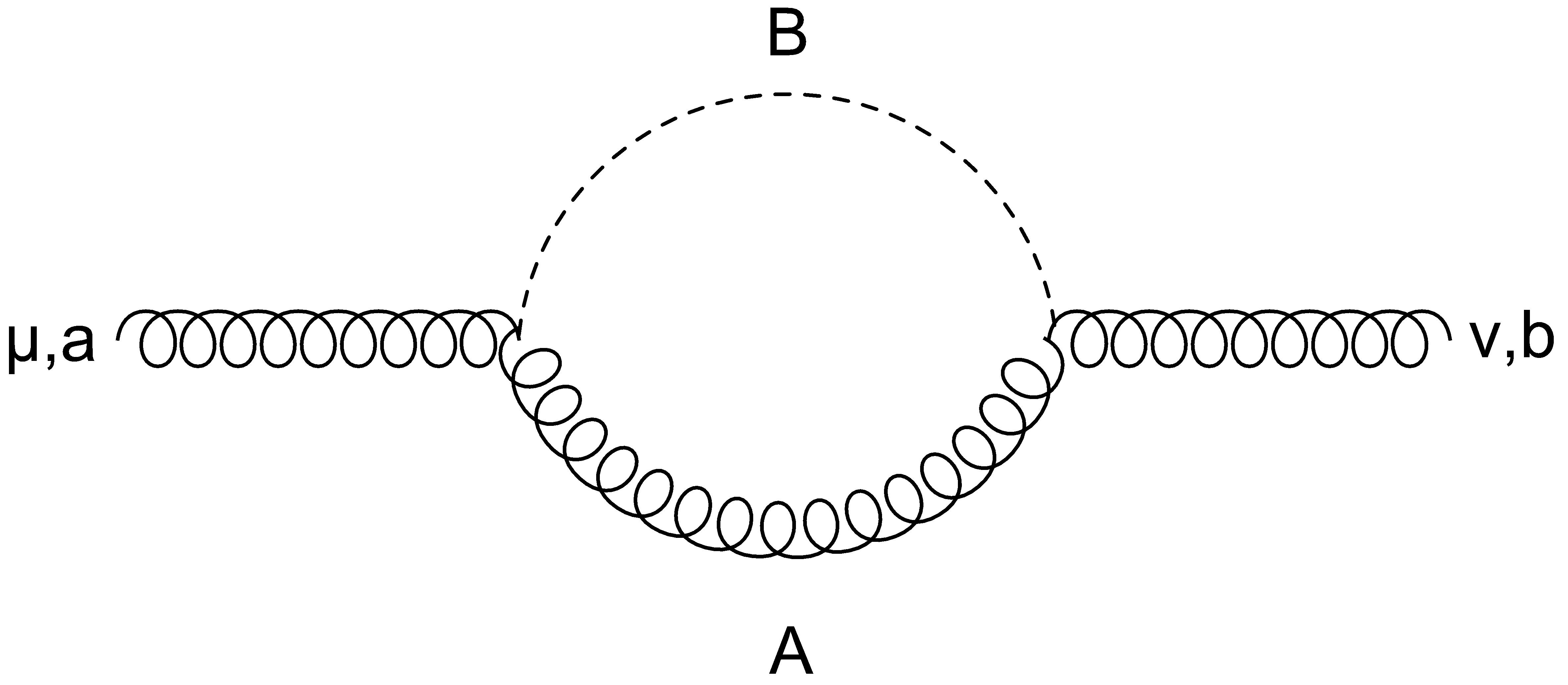} 
\caption{Loop diagram contains the $AAB$ coupling.}
\label{looprelvnt2}
\end{center}
\end{figure}

\begin{figure}[h!]
  \begin{center}
    \includegraphics[scale=0.06=3]{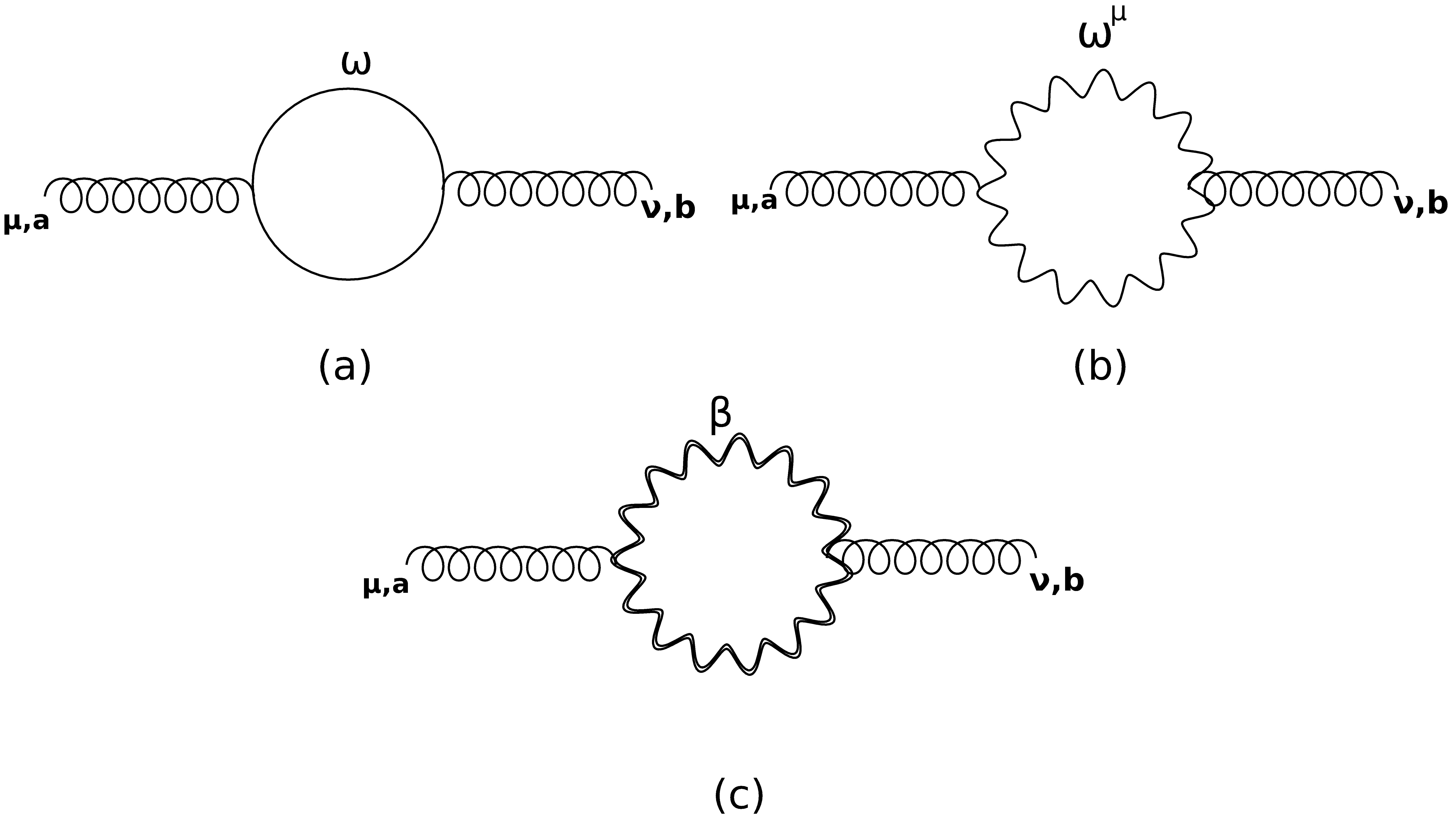} 
\caption{ Loops formed by (a) FP ghost of YM field, (b) vector ghost and (c) ghost of vector ghost.}
  \label{vecghst}
  \end{center}
\end{figure}

We easily reach  at the conclusion from Fig.~\ref{looprelvnt}a that the term $m^2/K^4$ in the propagator of massive 
YM field does not carry relevance under the approximation  considered here when $K$ is hard. Instead of propagator behaving as $\sim 1/k^2$, 
now we have to consider $1/(k^2-m^2)$. On the other hand,  the vertex rule of trilinear coupling among the massive gluon field 
is same as that of massless YM field. This makes the calculation easier. We have also noticed that the loop amplitude from  Fig.~\ref{looprelvnt}a in the HTL approximation at the leading order is same as that of the massless YM case  because of the structure of the propagator of massive YM field. On the other hand, the trilinear vertex rule among 
the massive YM field and its massless ghosts is same as that of the massless YM theory. These similarities imply that the thermal 
loop amplitude for Fig.~\ref{vecghst}a is same as found that of massless YM theory. The contributions from 
Figs.~\ref{looprelvnt}a and ~\ref{looprelvnt}b are
\begin{eqnarray}
\Pi^{\ref{looprelvnt}a}_{\mu\nu} & = &  \frac{g^2N_c}{2}\sumint_{\vec{k}}
\frac{-\delta_{\mu\nu}\left[5p^2 - 2p \cdot k + 2k^2 \right] + (d + 4) p_{\mu} p_{\nu} 
- (4d - 2) k_\mu \, k_\nu}{(k^2 - m^2) \left[(p - k)^2 - m^2\right]}, \label{conbt1} \nonumber\\
\Pi^{\ref{looprelvnt}b}_{\mu\nu} &=& - \frac{1}{2} g^2 \,N_c \delta_{\mu\nu}\sumint_{\vec{k}} \, \frac{2d}{(k^2 - m^2)}.
\label{conbt2}
\end{eqnarray}
Neglecting the terms  $\sim \mathcal{O}(p^2)$ in the numerator of integrand Eq.~(\ref{conbt1}), we get the spatial part as
\begin{eqnarray}
\Pi^{\ref{looprelvnt}a}_{ij} &\approx &  \frac{g^2\, N_c}{2}\sumint_{\vec{k}}
\frac{-\delta_{ij} \left[- 2 p \cdot k + 2k^2 \right] - (4d - 2)k_i \,k_j}{(k^2-m^2)\left[(p - k)^2 - m^2\right]} \nonumber \\
&=& -g^2 N_c\sumint_{\vec{k}}\frac{\delta_{ij}\left[- p \cdot k + k^2\right] + (2d -1)k_i\, k_j}{(k^2-m^2) \left[(p - k)^2 - m^2\right]}.
\label{loop7a}
\end{eqnarray}
Comparing  Eq.~(\ref{loop7a}) with Eq.~(\ref{genform}), we have $a_1=-1$, $a_3=0$,  $a_5=-5$ and 
$n=1$ for $d=3$.  Next, we consider the diagram in Fig.~\ref{looprelvnt}b, which provides the spatial part of loop amplitude as 
\begin{eqnarray}
\Pi^{\ref{looprelvnt}b}_{ij} &=&- g^2N_c \delta_{ij}\sumint_{\vec{k}}\frac{d}{(k^2-m^2)}
\approx - g^2d \, N_c \delta_{ij} \int_{\vec{k}} \frac{1}{2K}(1 + n_B(K)). \label{conbt22}
\end{eqnarray}
The loop amplitude corresponding to the diagram shown in Fig.~\ref{looprelvnt}c is given by\footnote{See the calculation of the numerator of the integrand in appendix B.}  
\begin{eqnarray}
\Pi^{\ref{looprelvnt}c}_{\mu\nu} = \frac{g^2N_c}{2}\sumint_{\vec{k}}\, \frac{2(d - 2)(k^2 - p \cdot k)\delta_{\mu \nu} 
+ (2d^2 - 3d + 4) \left[ 2k_\mu \, k_\nu -(p_\mu \, k_\nu + k_\mu \, p_\nu)\right]}{(k^2 - m^2)\{(p - k)^2 - m^2\}},
\end{eqnarray}
which  provides $a_1=2(d-2)$, $a_3=0$, $a_5=0$ and $n=2$. The  
loop amplitude from the loop diagram, shown in Fig.~\ref{looprelvnt}d is
\begin{eqnarray}
\Pi^{\ref{looprelvnt}d}_{ij} &=& \frac{1}{2}  g^2 N_c\delta_{ij} \left(d^2-3d+2\right)\sumint_{\vec{k}}\frac{1}{(k^2-m^2)} \nonumber\\
&\approx& \frac{1}{2} \,g^2 \left(d^2 - 3d + 2\right)\, N_c \delta_{ij} \int_{\vec{k}} \frac{1}{2K}\,(1 + n_B(K)).
\end{eqnarray}
Now, there is only one relevant loop diagram  involving $A$ and $B$ fields which left to consider
 is shown in Fig.~\ref{looprelvnt2}.
The loop amplitude corresponding to this diagram is obtained by neglecting the term $\sim {m^2}/{k^4}$ 
from the propagators of the field is given by
\begin{eqnarray}
\Pi^{\ref{looprelvnt2}}_{\mu\nu}& = &g^2 m^2N_c\sumint_{\vec{k}}\frac{(2d-2)\delta_{\mu\nu}}{(k^2-m^2)\left[(p-k)^2-m^2\right]},\\
\Rightarrow \Pi^{\ref{looprelvnt2}}_{ij}&=&2g^2 m^2N_c(d-1)\sumint_{\vec{k}}\frac{\delta_{ij}}{(k^2-m^2)\left[(p-k)^2-m^2\right]},
\end{eqnarray}
which gives in $a_3=2$, $a_1=a_5=0$ and $n=1$. Next we consider the ghost sector, which also contributes in the construction 
of HTL effective Lagrangian. The loop diagrams corresponding to fields, $\omega$, $\bar{\omega}$; $\omega^\mu$, $\bar{\omega}^\mu$ and $\beta$, $\bar{\beta}$ are shown in Fig.~\ref{vecghst}a, Fig.~\ref{vecghst}b and Fig.~\ref{vecghst}c respectively.

The loops are formed by the FP ghost of YM field in Fig.~\ref{vecghst}a, vector ghost in 
Fig.~\ref{vecghst}b  and ghost of the vector ghost in Fig.~\ref{vecghst}c. 
The loop amplitude from Fig.~\ref{vecghst}a is found as
\begin{eqnarray}
\Pi^{\ref{vecghst}a}_{\mu\nu} =  g^2 N_c \sumint_{\vec{k}}\frac{(k-p)_\mu \, k_\nu}{k^2(p-k)^2}
=  \frac{g^2N_c}{2}\sumint_{\vec{k}}\frac{2k_\mu \,k_\nu - p_\mu \, p_\nu}{k^2(p-k)^2},
\end{eqnarray}
where we have used the trick (cf. Eq.~(\ref{trick})) in the last step  because the loop integration contains 
the product of two identical propagators. Comparing with Eq.~(\ref{genform}), we see that $a_1=a_3=0$, $a_5=1$ and $n=1$.
Loop amplitude from the Fig.~\ref{vecghst}b is
\begin{eqnarray}
\Pi^{\ref{vecghst}b}_{\mu\nu} &=& -g^2 N_c\sumint_{\vec{k}}\frac{(d-2)p_\mu \,k_\nu - k_\mu \, k_\nu(d-1) + p_\nu \, k_\mu}{k^2 (p-k)^2} \nonumber\\
&=& -\frac{g^2N_c}{2}\sumint_{\vec{k}}\frac{(d-1)p_\mu \,p_\nu - 2k_\mu \, k_\nu(d-1)}{k^2 (p-k)^2}, \nonumber \\ 
\Rightarrow\Pi^{\ref{vecghst}b}_{ij} &\to & g^2N_c(d-1)\sumint_{\vec{k}}\frac{k_i\, k_j}{k^2 (p-k)^2}.
\end{eqnarray}
In the last step of the above integration, we have again used the same trick shown in Eq.~(\ref{trick}). In comparison
with Eq.~(\ref{genform}), we see that loop integration contributes to HTL effective Lagrangian with $a_1=a_3=0$ and $a_5=(d-1)$. 
The contribution from Fig.~\ref{vecghst}a is same as that of Fig.~\ref{vecghst}c, because of the similarity in 
the vertices of  the trilinear couplings $A\bar{\omega}\omega$ and $A\bar{\beta}\beta$. Hence 
adding up the contribution from the  ghost sectors we get,
\begin{eqnarray}
\Pi^{\ref{vecghst}a+\ref{vecghst}b+\ref{vecghst}c}_{ij}=  g^2 N_c \sumint_{\vec{k}}\frac{k_i\, k_j}{k^2 (p-k)^2}\left[2+(d-1)\right].
\end{eqnarray}
Comparing the generic expression in Eq.~(\ref{genform}) with the above equation, we get only $a_5=4$ when $d=3$ and $n=1$. Hence, we obtain the effective 
action from HTL approximation for topologically massive bosons in $d=3$ dimensions as
\begin{eqnarray}
\mathcal{L}_{eff} &=& - \frac{1}{4}F^{\mu\nu}_a \,F_{\mu\nu}^a+ m^2 \sumint_{\vec{k}}\left( A^\lambda(k)~A_{\lambda}(-k)- \frac{(k\cdot A(-k))(k\cdot A(k))}{k^2}\right)    \nonumber\\&&  +\, \int_K \tilde{m}^2(K) A^\mu(K) A_\mu (- K) + m_E^2\int d\Omega_v \left(\frac{1}{\mathcal{V} \cdot D}\mathcal{V}^\alpha F^a_{\alpha\mu}\right)
\left(\frac{1}{\mathcal{V} \cdot D}\mathcal{V}^\beta F_{a\beta}^\mu\right),
\label{effactn}
\end{eqnarray}
 where 
\begin{eqnarray}
\tilde{m}^2 &=& g^2 N_c\left(\frac{1}{2} \dfrac{T^2}{12} + 2 m^2 I \left(-1 + k^0 L(K)\right)\right),\label{mass1} \\  
m_E^2 &\approx & g^2 N_c \dfrac{T^2}{12}\label{mass2}.
\end{eqnarray}
In the final form of the effective action in Eq.~(\ref{effactn}), we have added the contribution obtained by integrating out the
$B$ field from the quadratic part of TMM action for the Lagrangian, given in Eq.~(\ref{nonab-bf}) (see Appendix C). 
The final form of the effective action in Eq.~(\ref{effactn}) also 
contains the  contributions from Fig.~\ref{looprelvnt}b and Fig.~\ref{looprelvnt}d. 
These contributions  are added to the coefficient $B$ in Eq.~(\ref{genform}) to provide the 
coefficient $\dfrac{1}{2}$  of $\dfrac{T^2}{12}$ in Eq.~(\ref{mass1}).


\section{Discussion}

We have constructed the HTL effective action for the topologically massive gauge theory. In the final form,  we have clearly shown how the Debye 
mass modified due to presence of bare mass of massive gauge bosons. The bare mass puts  an infrared cut-off in QCD at 
finite temperature. The infrared cut-off plays a crucial role in the perturbative analysis of transport coefficients, which are 
related to the response functions. These were believed to be in the non-perturbation regime in QCD at finite temperature. 
We have not considered any fermionic interaction with the massive YM gauge bosons. The fermions will have the same 
trilinear coupling with massive YM field as it has in massless YM theory. As a consequence, they provide same the 
contribution in the HTL approximated Lagrangian. There is no conserved local current constructed  from a trilinear 
coupling among fermions and $B_{\mu\nu}$ field. We have not calculated the transport coefficients from the HTL  
action for topologically massive gauge bosons when they are coupled with fermions. It will be very interesting to 
find the response functions from a matter coupled TMM at finite temperature.

We also see the other prospects of the TMM at finite temperature. In the massless YM theory at finite temperature, 
the phase transition can be explained by associating  with spontaneous broken symmetry. Massless YM field theory 
is invariant under $SU(N)/Z(N)$ group, where $Z(N)$ is centre of $SU(N)$ group. This symmetry  is  believed to be 
spontaneously broken at phase transition which is described by vacuum expectation value of Polyakov loop 
$\widetilde{L} = \dfrac{1}{N}\text{tr}\mathcal{P}\left(\exp i\oint_C A_0(\vec{x},t)\right)$, where $\mathcal{P}$ represents path 
ordering of the exponent and trace is taken to make $L$ invariant under $SU(N)$ symmetry. Taking the quarks to be static it can 
be shown that the implication of phase transition implies the spontaneous breaking of $SU(N)$ symmetry. But in TMM, 
there are massive gauge fields, which are in the adjoint representation of $SU(N)$ group. As a consequence, in the model, we have 
more general Polyakov loop 
\begin{eqnarray}
\widetilde{L}^{\text{gen}}\sim\text{tr}\mathcal{P}\left(\exp\left(i\oint A_0(\vec{x}, t) dx^0\oint_S B_{0i}dx^0 dx^i\right)\right),
\end{eqnarray}
where the closed path $C$ is loop and surface $S$ is taken in space-time. The physical significance and the 
behavior of  $\widetilde{L}^{\text{gen}}$ near the critical temperature  can be investigated thoroughly. It will be also interesting 
to consider thermal Bethe Salpeter equations from TMM. This may give the dynamics of the bound state massive gauge 
bosons at finite temperature.


\appendix
\renewcommand{\theequation}{A.\arabic{equation}}    
\setcounter{equation}{0}  


\section*{Appendix A: }
 
 
Here, we elaborate why $I$ (cf. Eq.~(\ref{divint})) appears divergent in HTL approximation. One of the integral representations of the modified Bessel function $K_1(ny)$ is defined as
\begin{eqnarray}
\displaystyle\int_0^\infty\frac{(x+y)e^{-\mu x}}{\sqrt{(x^2+2y x)}}~dx = y e^{y\mu}K_1(y\mu),\quad \left[\text{Re}~\mu>0, |\text{arg}~y|<\pi\right].
\end{eqnarray}
We take any one of the terms on r.h.s. of Eq.~(\ref{expan}).  Omitting some numerical factors, which hardly matter in the analysis, we consider the following integral 
\begin{eqnarray}
\int \frac{d^Dp}{(2\pi)^4}\frac{1}{E_1E_2}\frac{\left(n_B(E_1)-n_B(E_2)\right)}{ik_{0n}+E_2-E_1}.
\label{intgn}
\end{eqnarray}
Putting $E_1=\sqrt{P^2+m^2}$,  $E_2=\sqrt{(P-K)^2+m^2}$,  $n_B(E)=\displaystyle \sum_{s=1}^{\infty} e^{-s\beta E}$ 
and taking only one part (i.e. first term) of the integration
\begin{equation}
\mathcal{I}=\displaystyle\int\frac{d^DP}{(2\pi)^4}\frac{1}{\sqrt{p^2+m^2}\sqrt{(P-K)^2+m^2}}\sum_{s=1}^{\infty} \frac{e^{-s\beta E_1}}{ik_{0n}+\sqrt{(P-K)^2+m^2}-\sqrt{P^2+m^2}}.
\label{mainexpn}
\end{equation}
Neglecting $K$ with respect to $P$, $\mathcal{I}$ reduces  (in $D=3$) to the following form 
\begin{eqnarray}
\mathcal{I}\approx \frac{1}{ik_{0n}}\sum_{s=1}^{\infty} \mathcal{N} \int_0^\infty  e^{-s\beta \sqrt{P^2+m^2}} dP+\mathcal{T}(m^2),
\label{intgn2}
\end{eqnarray}
where the term $\mathcal{T}(m^2)$ is a
finite term multiplied by $m^2$. Changing the variable $P^2+m^2=y^2$ in the first term of Eq.~(\ref{intgn2}), we obtain
\begin{eqnarray}
\mathcal{I}=\frac{1}{ik_{0n}} \left(\sum_{s=1}^{\infty}\int_m^\infty \frac{y~e^{-s\beta y}}{\sqrt{y^2-m^2}}dy\right)+\mathcal{T}(m^2).
\end{eqnarray}
Now, again, redefining the variable as: $y-m=z$, we get
\begin{eqnarray}
\mathcal{I}&=&\frac{1}{ik_{0n}} \displaystyle\sum_{s=1}^{\infty}\left(\int_0^{\infty}\frac{(z+m) e^{-\beta s z}}{\sqrt{z(z+2m)}} dz 
+\mathcal{T}(m^2)\right)\nonumber \\ &=&\frac{1}{ik_{0n}} \displaystyle\sum_{s=1}^{\infty} m K_1(s\beta m)+ \mathcal{T}(m^2).
\end{eqnarray}
This is a convergent sum because of the behaviour of $K_1(s\beta m)$ in the limit $s\to \infty$ and $m\neq 0$, $s\neq 0$. 
However, the question of the divergence may be appeared again from the zero mode of $k_{0n}$, but here, again it is appeared due to 
approximation. It is very convenient to see the convergence of the integration in the lower and upper limits due to the presence 
of the non-zero poles of $P$ and finiteness of the exponential factor $e^{-\beta\sqrt{P^2+m^2}}$ of the integrand in Eq.~(\ref{mainexpn}). 
In summary, our purpose was to see how the integration over $p$ is convergent and this has been shown in a convenient way. 
Inclusion of the momentum $K$, in Eq.~(\ref{intgn2}), may change the nature of convergence but not the convergence. This 
inclusion also removes the appearance of the divergence due zero mode of $k_{0n}$ in the Mastubara sum.  This can be seen 
after neglecting the terms $\sim\mathcal{O}(K^2)$ from the integrand in Eq.~(\ref{mainexpn}). This yields
\begin{eqnarray}
\mathcal{I}\approx \displaystyle\sum_{s=1}^{\infty}m\left(  K_1(s\beta m)+2\sqrt{2}K_1(\sqrt{2}s \beta m)\right)L(K)+\cdots, 
\end{eqnarray}
 where $L(K)$ is defined in Eq.~(\ref{L(K)}). We have obtained the above expression only from one of the parts of $\mathcal{I}$ in Eq.~(\ref{mainexpn}). 
The same procedure can also work for its second part and we can obtain convergent terms containing modified Bessel functions.  This analysis clearly shows that the approximations made in Eq.~(\ref{expan1}) and Eq.~(\ref{expan2}) are responsible for the apprerant appearance of divergent $I$ in Eq.~(\ref{genform}).

\appendix
\renewcommand{\theequation}{B.\arabic{equation}}    
\setcounter{equation}{0}  


\section*{Appendix B: Calculation of amplitude of Fig.\ref{looprelvnt}c}
 

\begin{figure}[h!]
\begin{center}
\includegraphics[scale=0.06=3]{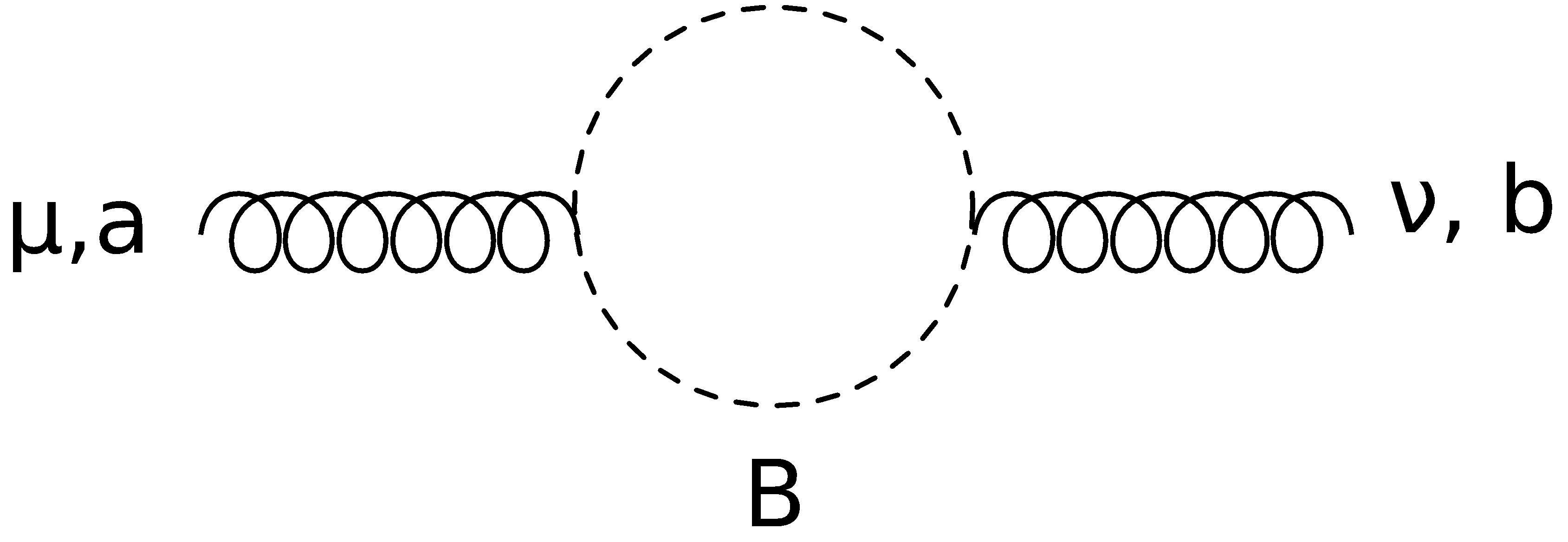} 
\caption{Loop diagram contains the $ABB$ coupling.}
\label{Bloop}
\end{center}
\end{figure}
The amplitude of the Fig.~\ref{Bloop} is given by
\begin{eqnarray}
\mathcal{M} &=& \frac{1}{16}\left[(2k-p)_\mu \, \eta_{\rho [ \alpha}\,\eta_{\beta ] \sigma} 
+ k _{[ \rho} \, \eta_{\sigma ][ \alpha}\,\eta_{\beta ] \mu} - \tilde{k} _{[ \alpha} \, \eta_{\beta ] 
[\rho}\,\eta_{\sigma ] \mu}\right] \left( \frac{\eta^{\alpha [ \alpha'}\,\eta^{\beta' ] \beta}}{k^2-m^2}
+\frac{m^2}{k^4}\frac{k^{[\alpha}k^{ [ \beta'}\,\eta^{\alpha' ] \beta]}}{k^2-m^2}\right) \times \nonumber\\
&& \left( \frac{\eta^{\rho [\rho'}\,\eta^{\sigma' ] \sigma}}{\tilde{k}^2 - m^2} 
+ \frac{m^2}{\tilde{k}^4}\frac{\tilde{k}^{[\rho}k^{ [ \sigma'}\,\eta^{\rho'] \sigma]}}{\tilde{k}^2 - m^2}\right)
\Big[(2k - p)_\nu \, \eta_{\alpha' [\rho'}\,\eta_{\sigma'] \beta'} + k _{[ \rho'} \, \eta_{\sigma' ] [ \alpha'}\,\eta_{\beta'] \nu} 
- \tilde{k}_{[ \alpha'} \, \eta_{\beta'] [\rho'}\,\eta_{\sigma'] \nu}\Big], \nonumber\\
\end{eqnarray}
where $\tilde{k}$ is given by $$\tilde{k} = p - k.$$
Now we will use the following property in the amplitude
\begin{eqnarray}
A^{[\mu \nu]}\, \eta_{\alpha [ \mu}\,\eta_{\nu ] \beta} = 2 A^{\mu \nu}\, \eta_{\alpha [ \mu}\,\eta_{\nu ] \beta},
\end{eqnarray}
so that we get
\begin{eqnarray}
\mathcal{M} &=& \frac{1}{16}\left[2(2k-p)_\mu \, \eta_{\rho  \alpha}\,\eta_{\beta  \sigma} 
+ 2k _{[ \rho} \, \eta_{\sigma ]  \alpha}\,\eta_{\beta  \mu} 
- 2\tilde{k} _{ \alpha} \, \eta_{\beta  [ \rho}\,\eta_{\sigma] \mu}\right] 
\left( \frac{\eta^{\alpha [ \alpha'}\,\eta^{\beta' ] \beta}}{k^2-m^2}
+ \frac{m^2}{k^4} \, \frac{k^{[\alpha}k^{[\beta'}\,\eta^{\alpha'] \beta]}}{k^2-m^2}\right) \times \nonumber\\
&& \left( \frac{\eta^{\rho [ \rho'}\,\eta^{\sigma' ] \sigma}}{\tilde{k}^2-m^2} 
+ \frac{m^2}{\tilde{k}^4} \frac{\tilde{k}^{[\rho}k^{[\sigma'}\,\eta^{\rho'] \sigma]}}{\tilde{k}^2 - m^2}\right)
\left[ 2(2k - p)_\nu \, \eta_{\alpha'  \rho'}\,\eta_{\sigma'  \beta'} + 2k _{ \rho'} \, \eta_{\sigma'  [\alpha'}\,\eta_{\beta' ] \nu} 
- 2\tilde{k} _{[ \alpha'} \, \eta_{\beta' ]  \rho'}\,\eta_{\sigma'  \nu}\right] \nonumber\\
&=& \frac{4}{16}\left[ (2k-p)_\mu \, \eta_{\rho  \alpha}\,\eta_{\beta  \sigma} 
+ k _{[ \rho} \, \eta_{\sigma ]  \alpha}\,\eta_{\beta  \mu} - \tilde{k} _{ \alpha} \, \eta_{\beta  [ \rho}\,\eta_{\sigma ] \mu}\right]
\left( \frac{\eta^{\alpha [ \alpha'}\,\eta^{\beta' ] \beta}}{k^2-m^2} 
+ \frac{m^2}{k^4}\frac{k^{[\alpha}k^{ [ \beta'}\,\eta^{\alpha' ] \beta]}}{k^2-m^2}\right) \times \nonumber\\
&& \left( \frac{\eta^{\rho [ \rho'}\,\eta^{\sigma' ] \sigma}}{\tilde{k}^2-m^2} 
+ \frac{m^2}{\tilde{k}^4}\frac{\tilde{k}^{[\rho}k^{ [ \sigma'}\,\eta^{\rho' ] \sigma]}}{\tilde{k}^2-m^2}\right)
\left[ (2k-p)_\nu \, \eta_{\alpha'  \rho'}\,\eta_{\sigma'  \beta'} + k_{ \rho'} \, \eta_{\sigma' [\alpha'}\,\eta_{\beta' ] \nu} 
- \tilde{k} _{[ \alpha'} \, \eta_{\beta' ]  \rho'}\,\eta_{\sigma'  \nu}\right] \nonumber\\
&=& \frac{4}{16}\left[ (2k-p)_\mu \, \eta_{\rho  \alpha}\,\eta_{\beta  \sigma} 
+ 2k _{ \rho} \, \eta_{\sigma   \alpha}\,\eta_{\beta  \mu} - 2\tilde{k} _{ \alpha} \, \eta_{\beta   \rho}\,\eta_{\sigma  \mu}\right] 
\left( \frac{\eta^{\alpha [ \alpha'}\,\eta^{\beta' ] \beta}}{k^2 - m^2}
+\frac{m^2}{k^4}\frac{k^{[\alpha}k^{ [ \beta'}\,\eta^{\alpha' ] \beta]}}{k^2-m^2}\right) \times \nonumber\\
&&\left( \frac{\eta^{\rho [ \rho'}\,\eta^{\sigma' ] \sigma}}{\tilde{k}^2 - m^2} 
+ \frac{m^2}{\tilde{k}^4}\frac{\tilde{k}^{[\rho}k^{ [ \sigma'}\,\eta^{\rho' ] \sigma]}}{\tilde{k}^2 - m^2}\right)
\left[ (2k-p)_\nu \, \eta_{\alpha'  \rho'}\,\eta_{\sigma'  \beta'} +2k _{ \rho'} \, \eta_{\sigma'   \alpha'}\,\eta_{\beta'  \nu} 
- 2\tilde{k} _{ \alpha'} \, \eta_{\beta'   \rho'}\,\eta_{\sigma'  \nu}\right]. \nonumber\\
\end{eqnarray} 
Now we ignore $O(m^2/k^4)$ and $O(m^2/k^6)$ terms to get,
\begin{eqnarray}
\mathcal{M} &=& \frac{4}{16(k^2-m^2)(\tilde{k}^2-m^2)}\left[(2k - p)_\mu \, \eta_{\rho  \alpha}\,\eta_{\beta  \sigma} 
+ 2k _{ \rho} \, \eta_{\sigma   \alpha}\,\eta_{\beta  \mu} - 2\tilde{k} _{ \alpha} \, \eta_{\beta   \rho}\,\eta_{\sigma  \mu}\right] 
\left(\eta^{\alpha [ \alpha'}\,\eta^{\beta' ] \beta}\right) \times \nonumber\\
&& \left(\eta^{\rho [ \rho'}\,\eta^{\sigma' ] \sigma}\right)\left[(2k - p)_\nu \, \eta_{\alpha'  \rho'}\,\eta_{\sigma'  \beta'} 
+ 2k_{ \rho'} \, \eta_{\sigma'   \alpha'}\,\eta_{\beta'  \nu} - 2\tilde{k} _{ \alpha'} \, \eta_{\beta'   \rho'}\,\eta_{\sigma'  \nu}\right] \nonumber\\
&=& \frac{4}{16(k^2 - m^2)\,(\tilde{k}^2-m^2)} \left[(2k - p)_\mu \, \delta^{[ \alpha'}_\rho \,\delta^{\beta' ]}_\sigma 
+ 2k _{ \rho} \, \delta^{[ \alpha'}_\sigma \,\delta^{\beta' ]}_\mu - 2\tilde{k} ^{ [\alpha'} \, \delta^{\beta']}_\rho\,\eta_{\sigma  \mu}\right] \times \nonumber\\
&& \left[ (2k-p)_\nu \, \delta^{[ \rho}_{\alpha'}\,\delta^{\sigma ]}_{\beta'} 
+2k ^{ [\rho} \, \delta^{\sigma]}_{\alpha'}\,\eta_{\beta'  \nu} - 2\tilde{k} _{ \alpha'} \, \delta^{[ \rho}_{\beta'}\,\delta^{\sigma ]}_{\nu}\right]. 
\end{eqnarray}
Let us denote the first and second square brackets by I and II respectively \emph{i.e.}
\begin{eqnarray}
\text{I} &=&  \left[ (2k-p)_\mu \, \delta^{[ \alpha'}_\rho \,\delta^{\beta' ]}_\sigma 
+ 2k_{ \rho} \, \delta^{[ \alpha'}_\sigma \,\delta^{\beta' ]}_\mu - 2\tilde{k} ^{ [\alpha'} \,\delta^{\beta']}_\rho\,\eta_{\sigma  \mu}\right], \\
\text{II} &=& \left[ (2k-p)_\nu \, \delta^{[ \rho}_{\alpha'}\,\delta^{\sigma ]}_{\beta'} 
+ 2k ^{ [\rho} \, \delta^{\sigma]}_{\alpha'}\,\eta_{\beta'  \nu} - 2\tilde{k} _{ \alpha'} \, \delta^{[ \rho}_{\beta'}\,\delta^{\sigma ]}_{\nu}\right]. 
\end{eqnarray}
The first term in I and first term in II are antisymmetric wrt $\alpha'$ and $\beta'$, so that we have
\begin{eqnarray}
\text{I}\times \text{II} &=& \left[ 2(2k-p)_\mu \, \delta^{ \alpha'}_\rho \,\delta^{\beta' }_\sigma 
+ 2k _{ \rho} \, \delta^{[ \alpha'}_\sigma \,\delta^{\beta' ]}_\mu - 2\tilde{k} ^{ [\alpha'} \, \delta^{\beta']}_\rho\,\eta_{\sigma  \mu}\right] \times \nonumber\\
&& \hskip 1cm \left[ (2k-p)_\nu \, \delta^{[ \rho}_{\alpha'}\,\delta^{\sigma ]}_{\beta'} 
+ 2k ^{ [\rho} \, \delta^{\sigma]}_{\alpha'}\,\eta_{\beta'  \nu} - 2\tilde{k} _{ \alpha'} \,\delta^{[ \rho}_{\beta'}\,\delta^{\sigma ]}_{\nu}\right].
\end{eqnarray}
Again the first term in first square bracket and first term in second square bracket above are 
antisymmetric w.r.t. $\rho$ and $\sigma$, so that we have
\begin{eqnarray}
\text{I}\times \text{II} &=& \left[2(2k-p)_\mu \,\delta^{ \alpha'}_\rho \,\delta^{\beta' }_\sigma 
+ 2k_{\rho} \, \delta^{[ \alpha'}_\sigma \,\delta^{\beta' ]}_\mu - 2\tilde{k} ^{ [\alpha'} \, \delta^{\beta']}_\rho\,\eta_{\sigma  \mu}\right] \times \nonumber\\
&& \hskip 1cm \left[ 2(2k-p)_\nu \, \delta^{ \rho}_{\alpha'}\,\delta^{\sigma }_{\beta'} 
+ 2k ^{ [\rho} \, \delta^{\sigma]}_{\alpha'}\, \eta_{\beta'  \nu} - 2\tilde{k} _{ \alpha'} \, \delta^{[ \rho}_{\beta'}\,\delta^{\sigma ]}_{\nu}\right],
\end{eqnarray} 
which implies
\begin{eqnarray}
\text{I}\times \text{II} &=& 4\left[(2k-p)_\mu \, \delta^{ \alpha'}_\rho \,\delta^{\beta' }_\sigma 
+ k _{ \rho} \, \delta^{[\alpha'}_\sigma \,\delta^{\beta' ]}_\mu - \tilde{k} ^{ [\alpha'} \, \delta^{\beta']}_\rho\,\eta_{\sigma  \mu}\right] \times \nonumber\\
&& \hskip 1cm \left[(2k - p)_\nu \, \delta^{ \rho}_{\alpha'}\,\delta^{\sigma }_{\beta'} 
+ k ^{[\rho} \, \delta^{\sigma]}_{\alpha'}\,\eta_{\beta'  \nu} - \tilde{k} _{ \alpha'} \, \delta^{[ \rho}_{\beta'}\,\delta^{\sigma ]}_{\nu}\right]. 
\end{eqnarray}  
Hence, the amplitude becomes
\begin{eqnarray}
\mathcal{M} &=& \frac{1}{(k^2-m^2)(\tilde{k}^2-m^2)}\left[ (2k-p)_\mu \, \delta^{ \alpha'}_\rho \,\delta^{\beta' }_\sigma 
+ k _{ \rho} \, \delta^{[ \alpha'}_\sigma \,\delta^{\beta' ]}_\mu - \tilde{k} ^{ [\alpha'} \, \delta^{\beta']}_\rho\,\eta_{\sigma  \mu}\right] \times \nonumber\\
&& \hskip 1cm \left[ (2k-p)_\nu \, \delta^{ \rho}_{\alpha'}\,\delta^{\sigma }_{\beta'} 
+ k ^{ [\rho} \, \delta^{\sigma]}_{\alpha'}\,\eta_{\beta'  \nu} - \tilde{k} _{ \alpha'} \, \delta^{[ \rho}_{\beta'}\,\delta^{\sigma ]}_{\nu}\right].
\end{eqnarray}
Again, let
\begin{eqnarray}
\text{I} &=&  \left[(2k-p)_\mu \, \delta^{ \alpha'}_\rho \,\delta^{\beta' }_\sigma 
+ k _{ \rho} \, \delta^{[ \alpha'}_\sigma \,\delta^{\beta' ]}_\mu - \tilde{k} ^{ [\alpha'} \, \delta^{\beta']}_\rho\,\eta_{\sigma  \mu}\right], \\
\text{II} &=& \left[ (2k-p)_\nu \, \delta^{ \rho}_{\alpha'}\,\delta^{\sigma }_{\beta'} 
+ k ^{ [\rho} \, \delta^{\sigma]}_{\alpha'}\,\eta_{\beta'  \nu} - \tilde{k} _{ \alpha'} \, \delta^{[ \rho}_{\beta'}\,\delta^{\sigma ]}_{\nu}\right].
\end{eqnarray} 
Also let $\text{I}_i$ denote the $i^{th}$ term in I, so that we have
\begin{eqnarray}
\text{I} &=& \text{I}_1 + \text{I}_2 -\text{I}_3, \\
\text{II} &=& \text{II}_1 + \text{II}_2 -\text{II}_3,   
 \end{eqnarray}
and
\begin{eqnarray}
\text{I}_1 \times \text{II}_1 &=& (2k-p)_\mu \, \delta^{ \alpha'}_\rho \,\delta^{\beta' }_\sigma 
\times (2k - p)_\nu \, \delta^{ \rho}_{\alpha'}\,\delta^{\sigma }_{\beta'} \nonumber\\
&=& d^2 \, (2k-p)_\mu  \, (2k-p)_\nu \nonumber\\
&\approx& d^2 \, \left[ 4k_\mu k_\nu -2 (k_\mu p_\nu + k_\nu p_\mu) \right],
\end{eqnarray}
\begin{eqnarray}
\text{I}_1 \times \text{II}_2 &=& (2k-p)_\mu \, \delta^{ \alpha'}_\rho \,\delta^{\beta' }_\sigma 
\times k ^{ [\rho} \, \delta^{\sigma]}_{\alpha'}\,\eta_{\beta'  \nu} \nonumber\\
&=& (2k-p)_\mu  \, \times k ^{ [\alpha'} \, \delta^{\sigma]}_{\alpha'}\,\eta_{\sigma  \nu} \nonumber\\
&=& (2k-p)_\mu  \, \times \left( k ^{ \alpha'} \, \delta^{\sigma}_{\alpha'} - k ^{ \sigma} \, \delta^{\alpha'}_{\alpha'}\right)\,\eta_{\sigma  \nu} \nonumber\\
&=& (1-d)\, (2k_\mu \, k_\nu-p_\mu \, k_\nu),
\end{eqnarray} 
\begin{eqnarray}
\text{I}_1 \times \text{II}_3 &=& (2k-p)_\mu \, \delta^{ \alpha'}_\rho \,\delta^{\beta' }_\sigma 
\times \tilde{k} _{ \alpha'} \, \delta^{[ \rho}_{\beta'}\,\delta^{\sigma ]}_{\nu} \nonumber\\
&=& (2k-p)_\mu \, \tilde{k} _{ \rho} \, \delta^{[ \rho}_{\sigma}\,\delta^{\sigma ]}_{\nu} \nonumber\\
&=& (2k-p)_\mu  \, \tilde{k} _{ \rho} \left( \delta^{\rho}_{\sigma}\, \delta^{\sigma}_{\nu} - \delta^{\sigma}_{\sigma}\, \delta^{\rho}_{\nu}\right)\nonumber\\
&=& (1-d)\, (2k_\mu \, \tilde{k}_\nu-p_\mu \, \tilde{k}_\nu) \nonumber\\
&\approx& (1-d)\, (2k_\mu \, p_\nu - 2k_\mu \, k_\nu+ p_\mu \, k_\nu),
\end{eqnarray}  
\begin{eqnarray}
\text{I}_2 \times \text{II}_1 &=& k _{ \rho} \, \delta^{[\alpha'}_\sigma \,\delta^{\beta']}_\mu 
\times (2k-p)_\nu \, \delta^{\rho}_{\alpha'}\,\delta^{\sigma}_{\beta'} \nonumber\\
&=& k_{ \alpha'} \, \delta^{[\alpha'}_{\beta'} \,\delta^{\beta' ]}_\mu \times (2k-p)_\nu  \nonumber\\
&=& k_{ \alpha'} \, (2k - p)_\nu  \left(\delta^{\alpha'}_{\beta'}\, \delta^{\beta'}_{\mu} - \delta^{\beta'}_{\beta'}\, \delta^{\alpha'}_{\mu}\right)\nonumber\\
&=& (1 - d) \, k_\mu \,(2k - p)_\nu  \nonumber\\
&=& (1 - d) \, \left[2k_\mu \, k_\nu - k_\mu \, p_\nu \right],
\end{eqnarray}
\begin{eqnarray}
\text{I}_2 \times \text{II}_2 &=& k _{ \rho} \, \delta^{[ \alpha'}_\sigma \,\delta^{\beta' ]}_\mu 
\times k^{ [\rho} \, \delta^{\sigma]}_{\alpha'}\,\eta_{\beta'  \nu} \nonumber\\
&=& k_{ \rho} \left(\delta^{\alpha'}_{\sigma}\, \delta^{\beta'}_{\mu} - \delta^{\beta'}_{\sigma}\, \delta^{\alpha'}_{\mu} \right) 
\left( k^\rho \delta^{\sigma}_{\alpha'} - k^\sigma \delta^{\rho}_{\alpha'}\right)\,\eta_{\beta'  \nu} \nonumber\\
&=& \left(\delta^{\alpha'}_{\sigma}\, \eta_{\mu\nu} - \eta_{\sigma \nu}\, \delta^{\alpha'}_{\mu} \right) 
\left( k^2 \delta^{\sigma}_{\alpha'} - k^{\sigma}k_{\alpha'}\right) \nonumber\\
&=& (d-2)\, k^2 \, \eta_{\mu \nu} + k_\mu k_\nu,
\end{eqnarray} 
\begin{eqnarray}
\text{I}_2 \times \text{II}_3 &=& k_{ \rho} \, \delta^{[ \alpha'}_\sigma \,\delta^{\beta' ]}_\mu 
\times \tilde{k} _{ \alpha'} \, \delta^{[ \rho}_{\beta'}\,\delta^{\sigma ]}_{\nu} \nonumber\\
&=& k_{ \rho} \, \tilde{k} _{ \alpha'} \, \left(\delta^{\alpha'}_{\sigma}\, \delta^{\beta'}_{\mu} - \delta^{\beta'}_{\sigma}\, \delta^{\alpha'}_{\mu} \right)
\left(\delta^{\rho}_{\beta'}\, \delta^{\sigma}_{\nu} - \delta^{\sigma}_{\beta'}\, \delta^{\rho}_{\nu} \right) \nonumber\\
&=& \left( \tilde{k}_{\sigma}\, \delta^{\beta'}_{\mu} - \tilde{k}_\mu \, \delta^{\beta'}_{\sigma} \right)
\left( k_{\beta'}\, \delta^{\sigma}_{\nu} - k_\nu\, \delta^{\sigma}_{\beta'} \right) \nonumber\\
&=& (d - 2)\tilde{k}_\mu \, k_\nu + \tilde{k}_\nu \, k_\mu \nonumber\\
&=& (d - 2) p_\mu \, k_\nu + p_\nu \, k_\mu - (d-1) k_\mu \, k_\nu,
\end{eqnarray}  
\begin{eqnarray}
\text{I}_3 \times \text{II}_1 &=& \tilde{k} ^{ [\alpha'} \, \delta^{\beta']}_\rho\,\eta_{\sigma  \mu} 
\times (2k-p)_\nu \, \delta^{ \rho}_{\alpha'}\,\delta^{\sigma }_{\beta'} \nonumber\\
&=&  (2k-p)_\nu \tilde{k} ^{ [\rho} \, \delta^{\beta']}_\rho\,\eta_{\beta'  \mu}   \nonumber\\
&=& (1-d)\, \tilde{k} _{ \mu}\, (2k -p)_\nu \nonumber\\
&\approx& (1-d)\left[ 2p_\mu\, k_\nu - 2k_\mu \, k_\nu + k_\mu \, p_\nu\right],
\end{eqnarray}
\begin{eqnarray}
\text{I}_3 \times \text{II}_2 &=& \tilde{k} ^{ [\alpha'} \, \delta^{\beta']}_\rho\,\eta_{\sigma  \mu} 
\times k ^{ [\rho} \, \delta^{\sigma]}_{\alpha'}\,\eta_{\beta'  \nu} \nonumber\\
&=& \left( \tilde{k}^{\alpha'} \delta^{\beta'}_{\rho} - \tilde{k}^{\beta'} \delta^{\alpha'}_{\rho}\right)
\left( k^\rho \delta^{\sigma}_{\alpha'} - k^\sigma \delta^{\rho}_{\alpha'}\right) \,\eta_{\sigma  \mu} \, \eta_{\beta'  \nu}  \nonumber\\
&=& \left( \tilde{k}^{\alpha'} \eta_{\rho \nu} - \tilde{k}_\nu \delta^{\alpha'}_{\rho}\right)
\left( k^\rho \eta_{\alpha' \mu} - k_\mu \delta^{\rho}_{\alpha'}\right) \nonumber\\
&=& \tilde{k}_\mu \, k_\nu + (d-2) \tilde{k}_\nu \, k_\mu \nonumber\\
&=& p_\mu \, k_\nu +(d-2)p_\nu \, k_\mu +(1-d)k_\mu \, k_\nu,
\end{eqnarray} 
\begin{eqnarray}
\text{I}_3 \times \text{II}_3 &=& \tilde{k}^{ [\alpha'} \, \delta^{\beta']}_\rho \,\eta_{\sigma\mu} 
\times \tilde{k}_{ \alpha'} \, \delta^{[ \rho}_{\beta'}\,\delta^{\sigma ]}_{\nu} \nonumber\\
&=& \left(\tilde{k}^{\alpha'} \delta^{\beta'}_{\rho} - \tilde{k}^{\beta'} \delta^{\alpha'}_{\rho}\right)
\left( \delta^{\rho}_{\beta'}\, \delta^{\sigma}_{\nu} - \delta^{\sigma}_{\beta'}\, \delta^{\rho}_{\nu} \right) \,\eta_{\sigma  \mu} \, \tilde{k}_{\alpha'}\nonumber\\
&=& \left( \tilde{k}^2 \delta^{\beta'}_{\rho} - \tilde{k}^{\beta'} \tilde{k}_{\rho}\right)
\left( \delta^{\rho}_{\beta'}\, \eta_{\mu \nu} - \eta_{\beta' \mu}\, \delta^{\rho}_{\nu} \right)\nonumber\\
&=& (d-2)\tilde{k}^2\, \eta_{\mu \nu}+\tilde{k}_\mu \, \tilde{k}_\nu \nonumber\\
&\approx & (d-2)(k^2-2 \, p \cdot k)\, \eta_{\mu \nu} + k_\mu \, k_\nu - (p_\mu \, k_\nu + k_\mu \, p_\nu).
\end{eqnarray} 
The amplitude is finally given by
\begin{eqnarray}
\mathcal{M} &=& \text{I}_1 \times \text{II}_1 + \text{I}_1 \times \text{II}_2 - \text{I}_1 \times \text{II}_3 
+ \text{I}_2 \times \text{II}_1 + \text{I}_2 \times \text{II}_2 - \text{I}_2 \times \text{II}_3 \nonumber\\
&-& \text{I}_3 \times \text{II}_1 - \text{I}_3 \times \text{II}_2 + \text{I}_3 \times \text{II}_3 \nonumber\\
&=& 2 (d - 2)(k^2 - p \cdot k) \eta_{\mu \nu} + (2d^2 - 3d + 4) \big[2k_\mu \, k_\nu -(p_\mu \, k_\nu + k_\mu \, p_\nu)\big].
\end{eqnarray}


\renewcommand{\theequation}{C.\arabic{equation}}    
\setcounter{equation}{0}  


\section*{Appendix C: `Integrating out' $B$ field at quadratic level }

We have already integrated out the hard modes of $B$ field from the non-Abelian TMM  by considering the tri-linear  and quartic 
interactions among $B$ and $A$ fields. But in the final form of the effective action in Eq.~(\ref{effactn}), we have to add classical action where 
$B$ field is integrated out its quadratic part. We consider the quadratic part:  
\begin{eqnarray}
\mathcal{L} = -\frac{1}{4}\,F^{\mu\nu} F_{\mu\nu} + \frac{1}{12}\, H^{\mu\nu\lambda} H_{\mu\nu\lambda}
+ \frac{m}{4}\,\varepsilon^{\mu\nu\rho\lambda} F_{\mu\nu} B_{\rho\lambda},
\label{abtmt}
\end{eqnarray}
where we have suppressed the gauge group indices. 
Introducing the gauge fixing term  
\begin{eqnarray}
\mathcal{L}_{GF}=\frac{1}{2\eta} \left(\partial_\mu B^{\mu\nu}\right)^2,
\end{eqnarray}
in the Lagrangian in Eq.~(\ref{abtmt}), where $\eta$ is gauge fixing parameters, we can find the 
two-point function of $B$ field. Hence, we can write the action corresponding to the above Lagrangian 
density in Eq.~(\ref{abtmt}) as
\begin{eqnarray}
\mathcal{S} &=& \int  d^4 x \Bigg (\frac{1}{2} \int d^4 y A_\mu(x)\left(\eta^{\mu\nu}\,\square - \partial^\mu \partial^\nu\right)\delta^4(x-y)A_\nu(y) \nonumber\\
&-& \frac{1}{4} \int d^4y \,B_{\mu\nu}(x) \,\Delta^{\mu\nu,\rho\lambda}(x,y, \eta) \,B_{\rho\lambda}(y) 
+ \frac{m}{2} \int d^4y \,j^{\mu\nu}(x)\, \delta^4(x-y) \,B_{\mu\nu}(y)\Bigg),
\label{actn}
\end{eqnarray}
where $\Delta(x, y, \eta)$ is the inverse of two-point function of $B$ field at the tree level and it has mass 
dimension $[\Delta]=6$ due to the inclusion of Dirac delta function. Here  
$j^{\alpha\beta} = \frac{1}{2}\, \varepsilon^{\alpha\beta\rho\lambda}F_{\rho\lambda}$. We can re-express the above expression as 
\begin{eqnarray}
\mathcal{S} =&-&\int d^4 x\, \frac{1}{4}\, F^{\mu\nu} F_{\mu\nu} \nonumber \\ 
&-& \frac{1}{4} \,\int d^4x \Bigg[\int d^4 y \left(B_{\mu\nu}(x) + \frac{m}{2}\int d^4z \,j^{\alpha\beta}(z) \, 
\Delta^{-1}_{\alpha\beta,\mu\nu}(z, x, \eta)\right) \times \nonumber\\ 
&& \Delta^{\mu\nu,\rho\lambda}(x,y, \eta)\left(B_{\rho\lambda}(y) + \frac{m}{2} \int d^4z \,\Delta^{-1}_{\rho\lambda, \alpha\beta}(y, z, \eta)\,
 j^{\alpha\beta}(z)\right)\Bigg] \nonumber\\ 
&+& \frac{m^2}{4} \int d^4k \, j^{\alpha\beta}(-k)~\Delta^{-1}_{\alpha\beta, \alpha'\beta'}(k)\, j^{\alpha'\beta'}(k).
\end{eqnarray}
 The appearance of the last term comes from the following steps:
\begin{eqnarray}
&& \frac{m^2}{16}\int d^4x~ d^4y ~d^4z~ d^4z'~ j^{\alpha\beta}(z)\,\Delta^{-1}_{\alpha\beta, \mu\nu}(z, x, \eta)\,
\Delta^{\mu\nu, \rho\lambda}(x, y, \eta) \, \Delta^{-1}_{\rho\lambda, \alpha'\beta'}(z', y, \eta)\, j^{\alpha'\beta'}(z')\nonumber\\ 
&& \qquad = \;\frac{m^2}{8}\int d^4y ~d^4z~ d^4z' \, j^{\alpha\beta}(z)~\delta^{\rho}_{[\alpha} \, \delta^\lambda_{\beta]} \, 
\delta^4(z-y)~\Delta^{-1}_{\rho\lambda, \alpha'\beta'}(z', y, \eta) \,  j^{\alpha'\beta'}(z') \label{2ndline} \nonumber\\ 
&& \qquad = \; \frac{m^2}{4}\int d^4z~ d^4z'\, j^{\alpha\beta}(z)~\Delta^{-1}_{\alpha\beta, \alpha'\beta'}(z, z', \eta)\, j^{\alpha'\beta'}(z') 
\label{eqn7}\nonumber \\ 
&& \qquad = \; \frac{m^2}{4}\int d^4k \,j^{\alpha\beta}(-k)~\Delta^{-1}_{\alpha\beta, \alpha'\beta'}(k)\, j^{\alpha'\beta'}(k),
\end{eqnarray}
where we have used 
\begin{eqnarray}
\frac{1}{2}\int d^4z \,\Delta^{-1}_{\rho\lambda, \alpha\beta}(x,z,\eta)\,\Delta^{\alpha\beta, \mu\nu}( z, y, \eta) 
= \delta^{\mu}_{[\rho}\, \delta^{\nu}_{\lambda]}\,\delta^4(x-y),
\end{eqnarray}
in the second line of  Eq.~(\ref{2ndline}). Using  $j^{\alpha\beta}= \frac{1}{2}\, \varepsilon^{\alpha\beta\rho\lambda}F_{\rho\lambda}= \varepsilon^{\alpha\beta\rho\lambda}\,\partial_\rho A_\lambda$, we can re-express the last term of the  Eq.~(\ref{actn}). Integrating by parts, we obtain
\begin{eqnarray}
&&\frac{m^2}{4}\int d^4z~ d^4z' j^{\alpha\beta}(z)~\Delta^{-1}_{\alpha\beta, \alpha'\beta'}(z, z', \eta) j^{\alpha'\beta'}(z'), \nonumber \\ 
&& \qquad = \; \frac{m^2}{4}\int d^4z~ d^4z' \varepsilon^{\rho\lambda\alpha\beta} A_\lambda(z)~\partial^z_\rho\, \partial^{z'}_{\rho'}\Delta^{-1}_{\alpha\beta, \alpha'\beta'}(z, z', \eta) ~\varepsilon^{\rho'\lambda'\alpha'\beta'}~A_{\lambda'}(z').
\end{eqnarray}
Then we can find
\begin{eqnarray}
&& \frac{m^2}{4}\, \varepsilon^{\rho\lambda\alpha\beta} ~\partial^z_\rho \, \partial^{z'}_{\rho'}\Delta^{-1}_{\alpha\beta, \alpha'\beta'}(z, z', \eta) 
~\varepsilon^{\rho'\lambda'\alpha'\beta} \nonumber\\ 
&& \qquad = \; \frac{m^2}{4} \,\varepsilon^{\rho\lambda\alpha\beta} ~ \int d^4k\, \Delta^{-1}_{\alpha\beta, \alpha'\beta'}(k)\, 
k_\rho\, k_{\rho'} \,e^{ik\cdot(z-z')} ~\varepsilon^{\rho'\lambda'\alpha'\beta'} \nonumber\\ 
&& \qquad = \; \frac{m^2}{4}\, \varepsilon^{\rho\lambda\alpha\beta} ~ \int d^4 k \,\frac{1}{k^2}\left(\eta_{\alpha[\alpha'} \, \eta_{\beta']\beta}
- (1 - \eta)\,\frac{k_{[\alpha} \, k_{[\alpha'} \, \eta_{\beta']\beta]}}{k^2}\right) 
k_\rho\, k_{\rho'} \,e^{ik\cdot(z-z')} ~\varepsilon^{\rho'\lambda'\alpha'\beta'} \nonumber\\
&& \qquad = \;  \frac{m^2}{4}\, \varepsilon^{\rho\lambda\alpha\beta} ~ \int d^4 k \, \frac{1}{k^2}~ 
\eta_{\alpha[\alpha'}\, \eta_{\beta']\beta}~ k_\rho \,k_{\rho'}\, e^{ik\cdot(z-z')} ~\varepsilon^{\rho'\lambda'\alpha'\beta'}\nonumber\\  
&& \qquad  =\; -\, m^2\int d^4 k \,\frac{1}{k^2}\left(k^2\, \eta^{\lambda \lambda'} - k^\lambda\, k^{\lambda'} \right)e^{ik\cdot(z-z')}.
\end{eqnarray}
As a consequence, the ``effective" action from the qudratic part at the tree level 
(where the degrees of freedom of $B$ field is integrated out) is given as follows 
\begin{eqnarray}
\mathcal{S}_{eff}&=& -\frac{1}{4}\,F^{\mu\nu} F_{\mu\nu}  + m^2 \int d^4z~ d^4z' A_\lambda(z)~\int d^4 k\frac{1}{k^2}\left(k^2 \, \eta^{\lambda \lambda'} 
- k^\lambda\,k^{\lambda'}\right)e^{ik\cdot(z-z')}~A_{\lambda'}(z') \nonumber\\ 
&=& - \frac{1}{4}\,F^{\mu\nu} F_{\mu\nu}  + m^2 \int d^4k~ A^\lambda(k)~A_{\lambda}(-k) - m^2 \int d^4k \,\frac{(k\cdot A(-k))(k\cdot A(k))}{k^2}.
\end{eqnarray}

\section*{Acknowledgements}
DM thanks Department of Atomic Energy, Govt. of India for financial support and RK would like to thank  
UGC, Government of India, New Delhi, for financial support under the PDFSS scheme. 
We would like to thank  Prof. S. Mrówczyński for bringing his recent work on the general form of 
HTL-effective action for massless gauge bosons~\cite{stanis} into our notice.

\end{document}